\documentclass[preprint, sort&compress]{elsarticle}

\usepackage{a4wide}
\usepackage{titlecaps}

\usepackage[hidelinks]{hyperref}
\usepackage{amsmath,amsfonts,amssymb}
\usepackage{enumitem}
\usepackage{graphicx}
\graphicspath{{./figures/}}
\usepackage[noabbrev]{cleveref}
\Crefformat{equation}{#2equation (#1)#3}
\Crefrangeformat{equation}{#2equations (#1)#3}
\crefformat{equation}{#2equation (#1)#3}
\crefrangeformat{equation}{#2equations (#1)#3}
\crefformat{appendix}{{#2Appendix #1#3}}
\usepackage{caption, subcaption}
\usepackage{multirow}
\usepackage{siunitx}
\usepackage[toc,page]{appendix}
\newcommand{\var}[1]{{\operatorname{\mathit{#1}}}}

\crefname{algocf}{algorithm}{algorithms}
\Crefname{algocf}{Algorithm}{Algorithms}

\usepackage{algpseudocode}
\usepackage[vlined, linesnumbered]{algorithm2e}
\makeatletter
\renewcommand{\@algocf@capt@plain}{above}
\newcommand{\nosemic}{\renewcommand{\@endalgocfline}{\relax}}
\newcommand{\dosemic}{\renewcommand{\@endalgocfline}{\algocf@endline}}
\newcommand{\pushline}{\Indp}
\newcommand{\popline}{\Indm\dosemic}
\makeatother

\journal{Journal of Sound and Vibration}

\begin{document}

\begin{frontmatter}

\title{Informative Bayesian Tools for Damage Localisation by Decomposition of Lamb Wave Signals}

\author[UoS_address]{Marcus Haywood-Alexander\corref{mycorrespondingauthor}}
\cortext[mycorrespondingauthor]{Corresponding author}
\ead{mhaywood-alexander1@sheffield.ac.uk}

\author[UoS_address]{Nikolaos Dervilis}
\author[UoS_address]{Keith Worden}
\author[Strath_address]{Gordon Dobie}
\author[UoS_address]{Timothy J.\ Rogers}

\address[UoS_address]{Dynamics Research Group, Department of Mechanical Engineering, Mappin Street, University of Sheffield, Sheffield, S1 3JD, UK}
\address[Strath_address]{Electronic and Electrical Engineering, Royal College Building, 204 George Street, Glasgow, G1 1XW, UK}

\begin{abstract}
Ultrasonic guided waves offer a convenient and practical approach to structural health monitoring and non-destructive evaluation, thanks to some distinct advantages. %
Guided waves, in particular Lamb waves, can be used to localise damage by utilising prior knowledge of propagation and reflection characteristics. %
Typical localisation methods make use of the time of arrival of waves emitted or reflected from the damage, the simplest of which involves triangulation (with a known wave speed). %
In order to obtain reflection information, it is useful to decompose the measured signal into the expected waves propagating directly from the actuation source in the absence of damage, called a \emph{baseline}, and for this paper referred to as \emph{nominal waves}. %
This decomposition allows for determination of the residual signal which contains only waves from reflection sources such as damage, boundaries or other local inhomogeneities. %
Previous decomposition methods make use of accurate analytical models, but there is a gap in methods of decomposition for complex materials and structures. %
A new method is shown here which uses a Bayesian approach to decompose single-source signals, requiring only prior information on surface displacement along the propagation path. %
This Bayesian decomposition has the advantage of generating a distribution of possible nominal signals and allows for quantification of the uncertainty of the expected signal. %
Furthermore, the approach produces inherent parametric features which correlate to known physics of guided waves, and likelihood estimates can be used to assess the quality of the decomposition. %
In this paper, the decomposition method is demonstrated on data from a simulation of guided wave propagation in a small aluminium plate, using the \emph{local interaction simulation approach}, for a damaged and undamaged case. %
Analysis of the decomposition method is done in three ways; inspect individual decomposed signals, track the inherently produced parametric features along propagation distance, and use method in a localisation strategy. %
The localisation method is demonstrated using the decomposed signal at several sensor locations and triangulates for the source of reflected waves from damage. %
The Bayesian decomposition was found to work well in returning signals containing only reflected waves, as well as obtaining parametric features that can be used to assess damage and confidence in the decomposed wave. %
The use of these waves in the localisation method returned estimates accurate to within 1mm in many sensor configurations. %
Leading on from the work shown here, the paper finishes with future work; the authors intend to extend this method to scenarios where less prior knowledge is available. %
\end{abstract}


\end{frontmatter}


\section{Introduction}

The use of ultrasonic guided waves (UGWs) for non-destructive evaluation (NDE) and structural health monitoring (SHM) strategies \cite{Rose2004} can offer a number of advantages, such as long range and accurate sizing potential, greater sensitivity and cost effectiveness. %
UGWs consist of two types of high-frequency stress waves: Rayleigh waves which propagate on a surface or, in plate-like media where the thickness is sufficiently small compared to the wavelength, as Lamb waves \cite{Viktorov1967, Worden2001, Rose2014}. %
A particular characteristic of Lamb waves is their separation into symmetric and antisymmetric modes, in the former of which the surfaces oscillate in opposite directions at equal propagation distance, whereas the latter oscillate in the same directions. %
Higher-order wave modes will also be present with an increased \emph{frequency-thickness} product. %
A \emph{wave-packet} is a single burst containing multiple wave modes of different frequencies and shapes; for Lamb waves these will contain both types of modes. %
The propagation velocity of Lamb waves depends on the central frequency of the wave and will vary between the modes present; therefore a wave-packet of mixed wavelengths will spread out in space, i.e., it will \emph{disperse.} %

As the wavelength of ultrasonic guided waves is small, using them for an SHM or NDE strategy would allow one to move up Rytter's hierarchy \cite{Rytter1993, farrar2012structural}, going from damage detection to assessing location or extent \cite{Kundu2019}. %
When Lamb waves interact with damage, they will reflect and scatter \cite{Lee2003}, causing additional waves to propagate from the damage location. %
Therefore, any wave captured by a sensor at a given location on a plate with damage, will contain information on the waves directly received from the actuation source, here named \emph{nominal waves}, and any reflections or conversions that are caused by the presence of damage. %
The received signal will be a superposition of the nominal waves -- of which there will be multiple modes -- and reflected waves. %
By decomposing this multi-mode signal into the nominal waves at three given propagation distances, a signal can be constructed which contains only the contents of the signal from reflections. %
The difference in arrival times of these reflections can be used, along with known wave velocities, to determine the location of the source of scattering. %

There has been evidence of useful localisation techniques that do not require decomposition, such as via the use of piezoelectric rosettes \cite{Kijanka2015}; however, the cost of using such hardware may quickly rise for large-scale systems. %
A novel method has been proposed by Rebillat \textit{et al}., involving the decomposition of three-way tensors constructed from `actuator', `sensor' and `time dimensions' that shows robustness \cite{Rebillat2020}. %
However, this method requires that data are collected from multiple actuation sources and sensor locations, so it may be advantageous in some systems to be able to localise using fewer such locations, thus reducing data usage and processing time. %
Methods that do not require decomposition of measured signals, such as the ones mentioned here, use either costly equipment or are computationally expensive; consequently, there is an opportunity for a computationally-efficient strategy involving simple equipment, to be developed. %

Some multi-mode decomposition techniques have been proposed with a variety of approaches. %
One such approach is a parameter-based iterative procedure to match the shape of the wave with the physical equations governing a Lamb wave in homogeneous materials \cite{Zhou2019}. %
Other methods involve the usage of a series of 1D and 2D bandpass filters on full-field laser scans over the area of wave propagation \cite{Tian2014}, identification of ratio features from signal processing of the received signal \cite{Park2014}, and using concentric ring and circular PZTs \cite{Yeum2011}. %
Many current individual-signal decomposition methods require accurate previous models of guided wave propagation in the material being inspected, analytical models of which are difficult to attain accurately for complex materials. %
Such an example would be for fibre-matrix composites, where the attenuation varies with respect to fibre-orientation \cite{Haywood2021}. %
Therefore, a method for single-source multi-mode decomposition was developed that only requires prior data on Lamb wave propagation in one direction \cite{Haywood2021SPIE}. %

Once any reflection signals are determined, the reflection from damage can then be treated as a new wave source; the time of the reflection event and distance of the damage from each sensor is unknown. %
After formulating the problem in this way, methods of localisation vary depending on the prior information known of the system \cite{Tobias1976, Kundu2014}. %
Common practice for localisation of unknown initiation time is relatively simple, involving the difference in time of arrival of the wave between sensors at known locations \cite{Tobias1976}. %
As numerical solutions of the Lamb waves give prior information on the wave velocity, given a frequency, the localisation step becomes a simple computation problem. %
However, for situations where the wave speed is unknown, localisation is still possible using an iterative optimisation procedure \cite{Kundu2014}. %
As one advantage of the decomposition strategy used here is its applicability for use with complex geometries and materials \cite{Haywood2021SPIE}, it is important to also consider localisation in situations where confidence in prior wave velocity is limited. %
For such situations, it may be useful to take a probabilistic approach, such as in \cite{flynn2011maximum} and \cite{jones2022bayesian}. %
Using such an approach would also allow for quantification of the uncertainty in location, an important metric to consider in NDE or SHM strategies. %

In this paper, a method is proposed which utilises prior knowledge of the nominal symmetric and antisymmetric modes to obtain reflection signals at a sensor location and use this to triangulate for the source of reflection. %
Lamb waves were simulated in a homogeneous plate using the local interaction simulation approach (LISA) \cite{Delsanto1997,Dobie2011}, for an undamaged plate and a damaged plate. %
In order to obtain the individual nominal signals of a Lamb wave as it propagates through space, first a full-field multi-mode separation technique was applied which utilises a forward-backward, two-dimensional Fourier transform (TDFT) method and dispersion curve information \cite{Alleyne1991,Haywood2020}. %
The simulation is then used to obtain several `sensor' signals at various locations over the plate, and are treated as \emph{single-source} signals.
These decomposed signals are then used to represent the expected nominal wave shape at a given location and are used as the basis for decomposing an individual signal into the nominal and \emph{residual} signals. %
The reflected waves are contained in the residual signal, determination of the onset of which gives the time of arrival of the reflection at the sensor. %
Various combinations of three single-source signals are used to triangulate the location of the reflection source. %
By using this methodology, it is shown here how damage location in a plate can be determined using a computationally-efficient strategy and simple equipment. %

A Bayesian approach is used for the single-source decomposition stage, which gives a distribution of solutions, and the mean is used as the solution. %
This probabilistic approach gives a number of advantages, which are listed and expanded upon below. %
\begin{itemize} 
	\item The predicted distribution of possible signals allows for improved analysis of residual signals, as the estimated uncertainty can be used for a number of sub-methodologies within the application. For example, the uncertainty can be used for direct assessment of the decomposition quality, or to implement an uncertainty limit over for decomposition use.  
	\item Uncertainty quantification is done at an early stage of the localisation procedure and can be propagated through to the determined location. This level of uncertainty can directly be implemented into the uncertainty of the time-of-arrival of the reflected wave, which is trivially propagated through to the uncertainty in location. %
	\item The results produce inherent parametric features which are indicative of energy-based features of UGWs \cite{Haywood2021SPIE}. Features are commonly used in all stages of NDE/SHM strategies, and so minimising cost by reducing the number of computational steps required is advantageous. %
\end{itemize}

The novelty of the method shown here in comparison to previous methods is twofold: the determination of a \emph{nominal wave dictionary} (NWD), and the probabilistic approach to the single-source decomposition. %
The nominal wave dictionary approach allows the method to be applied to materials where accurate analytical models of the wave signals are difficult to obtain. %
As the full-field reconstruction stage will reconstruct all reflection or converted signals, as long as they are included in the time width of the data passed into the 2DFT algorithm, any arbitrary complex material or geometry can be used with this method. %
Depending on the complexity of the material and geometry, formation of the NWD data can be done by either numerical methods, such as finite element modelling, or experimental methods, such as using a Scanning Laser Doppler Vibrometer. %
This is shown here where the influence of multi-mode signals on individual wave modes are included in the reconstructed signal. %
The strategy presented in this paper requires only surface displacement along a single propagation path and so can be readily used in analytical, numerical or experimental regimes. %
Modern health monitoring frameworks use probabilistic approaches for purposes such as novelty detection and uncertainty quantification \cite{farrar2012structural}. %
By using a methodology with a probabilistic approach, such as that proposed here, this lends itself to such health monitoring frameworks which aim to use, and propagate, probabilistic methods throughout. %
It is useful to quantify uncertainty in the signal to determine whether residual signals are from additional waves, rather than more complex phenomena such as continuous mode conversion \cite{Mook2014,Willberg2012}. %

The overall steps of the work presented here are as follows:
\begin{enumerate}[label=\alph*)]
	\item Using LISA, simulate propagation of a Lamb wave in a plate under damage-free conditions, giving a known wave field at all points on the surface.
	\item Reduce this response data to a subset containing only data along a single propagation direction and use this to determine a `dictionary' of individual wave modes at known propagation distances.
	\item Simulate propagation of a Lamb wave in an identical plate under a known damage condition.
	\item Extract from this full-field wave data a subset of single-source signals at known locations.
	\item Using a Bayesian linear regression (BLR) decomposition technique, determine the predicted nominal waves at the single-source signal location.
	\item Assess the quality of the decomposed signal and determine the residual signal which contains reflected waves.
	\item Use the time of arrival of reflected waves with known wave speeds to triangulate the location of the reflection source: i.e.\ the damage.
\end{enumerate}

This paper begins by introducing the key physics of ultrasonic guided waves, how they interact with damage, and the simulation technique used. %
Following this, the multi-stage localisation procedure is outlined in \Cref{sec:methodology} and each stage is described in further detail in subsections. %
\Cref{sec:decomp_results} shows the results of Bayesian decomposition of the damaged plate wave signals, and discusses the confidence of the decomposed signal, as well as the parametric features obtained.
In \Cref{sec:one_dim_results}, the results of the one-dimensional localisation are presented to demonstrate the concept of reflection-based localisation. %
The results of the two-dimensional localisation procedure using the full methodology for various sensor configurations are then shown in \Cref{sec:two_dim_results}. %
The paper finishes by discussing the results shown and the accuracy of the method, and concluding with further work the authors intend to do, which follows on from the results shown here. %

\section{Lamb wave propagation in plates}

Guided waves are used in many engineering applications, such as non-destructive evaluation (NDE) and structural health monitoring (SHM). %
In order to better apply these waves for such strategies, prior knowledge of their behaviour is essential. %
Such waves undergo an interesting phenomenon when they occur in particular structures, such as rods, hollow cylinders and plates; they propagate primarily in the longitudinal direction perpendicular to oscillation and are known as guided waves. %
When such waves are guided by propagation along the surface of a medium, they are called Rayleigh waves. %
However, if a wave travels in a bounded medium, between two surfaces, where the wavelength is sufficiently long compared to the distance between these surfaces, often exhibited in plates, it is called a Lamb wave. %
Overviews of the derivations and characteristics of such waves are well described by Viktorov \cite{Viktorov1967}, Worden \cite{Worden2001} and Rose \cite{Rose2014}. %

\subsection{Physics of Lamb waves}

\begin{figure}
    \centering
    \includegraphics[width=0.8\textwidth]{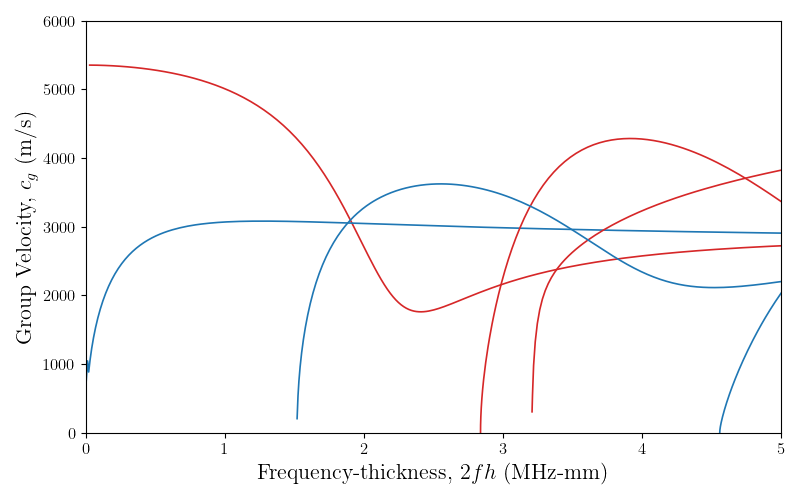}
    \caption{Dispersion curves for the group velocity of a 1mm thick aluminium plate, calculated using the DISPERSE software \cite{Pavlakovic1997disperse}. The blue lines indicate the antisymmetric $A_n$ modes, and the red lines indicate the symmetric $S_n$ modes, with the order they appear along the x-axis being the order $n$ of the mode. The aluminium plate is modelled with density of 2710 \si{kg/m^3}, and wave speeds $c_L$ = 6.42 \si{m/ms} \& $c_T$ = 3.04 \si{m/ms}.}
    \label{fig:disp_examp}
\end{figure}

Elastic waves in orthotropic, inhomogeneous media are described by the elastodynamic equation \cite{Achenbach1973},
\begin{equation}
   \partial_l(S_{klmn}w_{m,n}) = \rho\ddot{w}_k \qquad (k,l,m,n = 1,3)
   \label{eq:elastodynamic}
\end{equation}
where $S$ is the stiffness tensor, $\rho$ is the material density, $w$ is the displacement field for which the comma represents differentiation with respect to space and the double dot represents double differentiation with respect to time. %
In bounded media these waves will show as Lamb waves, which in isotropic elastic media will exhibit two distinct modes: symmetric and antisymmetric. %
There are two characteristic equations for Lamb waves which describe their behaviour in given media:
\begin{equation}
	\frac{\tan(qh)}{\tan(ph)} = - \frac{4k^2pq}{(q^2-k^2)^2}
	\label{eq:symMode}
\end{equation}
for the symmetric modes and,
\begin{equation}
	\frac{\tan(ph)}{\tan(qh)} = - \frac{4k^2pq}{(q^2-k^2)^2}
	\label{eq:antisymMode}
\end{equation}
for the antisymmetric modes. The definitions of $p$ and $q$ are given by,
\begin{equation}
	p = \frac{\omega}{\sqrt{c_L^2-c^2}}, \qquad q = \frac{\omega}{\sqrt{c_T^2-c}}
\end{equation}
where $2h$ is the plate thickness, $k$ is the real wavenumber, $\omega$ is the central frequency, $c$ is the bulk wave speed, $c_L$ is the longitudinal wave speed and $c_T$ is the transverse wave speed. %
Given a value of $\omega$, \cref{eq:symMode,eq:antisymMode} specify allowed values of $c$ for either mode; as $c$ is a function of $\omega$ this means that the waves are \emph{dispersive}. %
As $\omega$ only enters into \cref{eq:symMode,eq:antisymMode} as a product with $h$, dispersion curves are often presented in terms of the \emph{frequency-thickness product} (FTP). %
Solutions to these equations are determined numerically, and plots of the wave velocity with respect to frequency-thickness are called  \emph{dispersion curves}. %
Software packages are available to generate numerically-determined dispersion curve information -- an example being \emph{DISPERSE} \cite{Pavlakovic1997disperse} (see \Cref{fig:disp_examp}). %

Dispersion curves can also be determined from experimental regimes by the use of a \emph{two-dimensional Fourier transform} (TDFT); this is done by recording the surface displacement of a Lamb wave, spatially sampled along its propagation path, to generate the time-distance [$\var{t-x}$] space. %
Passing this through a TDFT then provides a transformation to the frequency-wavenumber [$\var{f-k}$] space \cite{Alleyne1991}.

Typical presentations of dispersion curves are as the relation of FTP to \emph{group velocity}, \emph{phase velocity} or \emph{wavenumber}. %
For this paper, the group velocity and wavenumber dispersion curves are important, as the group velocity is required for triangulation of reflection signals, and the wavenumber dispersion curve is the type generated by experiment.


\subsection{Effect of damage}

The interaction of Lamb waves with local inhomogeneities is well studied and the effects are well documented \cite{Alleyne1992}. %
Almost intuitively, when a Lamb interacts with damage, it will reflect and scatter in directions other than its original propagation direction collinear with the actuation source and the damage. %
The reflection of these waves when interacting with damage varies between modes, and is dependent on the relation between wavelength and damage size \cite{Lowe2002s0, Lowe2002a0}. %
A particular phenomenon that occurs when Lamb waves interact with damage is that of \emph{mode conversion} \cite{Alleyne1992}, where modes are converted into others as a result of variations in plate dimensions. %
This phenomenon is directly linked to the reflection characteristics of the interaction. %
In the rest of this paper, the incident waves that are expected to be observed at a location regardless of the presence of damage are referred to as the \emph{nominal waves}. %
At any given location away from the damage, signals received at a sensor will be a superposition of the nominal waves and any reflected or scattered waves resulting from the nominal wave interaction with the damage. %
In the received signal, these reflected waves will always be later in time than their respective nominal waves. %

\subsection{LISA Simulation}
\label{sec:LISA}

Guided waves were simulated for this work using the local interaction simulation approach (LISA), as this captures the multi-mode wave signals well. %
The LISA simulation method uses iterative difference equations and is based on a sharp-interface model. %
This allows LISA to incorporate the effects of boundaries and inhomogeneities with ease -- a primary benefit of using the method -- as well as faster computing time in comparison to finite element analysis (FEA). %
LISA is well documented \cite{Delsanto1997,Dobie2011,Nadella2013}, but a brief overview will be given here. %
A key difference between this method and standard finite difference (FD) approaches is that LISA solves a discrete form of \Cref{eq:elastodynamic} exactly, modelling physical phenomena without other approximations, whereas the FD is the solution of the partial differential equation after discretisation. %

A finite-difference formulation is used on \Cref{eq:elastodynamic} to generate iterative equations that can be applied for a given point in space. %
The derivations of these equations can be followed in \cite{Delsanto1997}, which is achieved by discretising the grid into evenly-sized cells and considering these cells as a series of springs and masses which populate the medium. %
At any point, which is neighboured by eight cells, the sharp-interface model is used to average the properties of the neighbouring cells. %
It is assumed at each point that the material properties and displacements are continuous, whereas interfaces of cells are treated as discontinuous. %
As the final iterative equations are quite lengthy and not the focus of this paper, they are not given here, but the reader can refer to the work by Delsanto \textit{et al.} \cite{Delsanto1997} and Sundararaman \& Adams \cite{Sundararaman2008}. %

\begin{table}[h!]
	\centering
	\begin{tabular}{| l | l |}
		\hline
		Material & Aluminium \\
		\hline
		Density, $\rho$ & 2710 \si{kg /m^3}\\
		\hline
		Longitudinal sound speed, $c_L$ & 6420 \si{m/s} \\
		\hline
		Transverse sound speed, $c_T$ & 3040 \si{m/s} \\
		\hline
		Width x Length & 300 x 300 \si{mm} \\
		\hline
		Plate thickness & 1 \si{mm}\\
		\hline
		Cell dimension, $\varepsilon$ & 0.25 \si{mm} \\
		\hline
		Time step, $\tau$ & 0.01496 \si{\micro\second} \\
		\hline
	\end{tabular}
	\caption{Properties of Lamb wave in plate simulation performed using LISA approach.}
	\label{tab:LISA_props}
\end{table}

\begin{figure}[!h]
	\centering
	\includegraphics[width = 0.7\textwidth]{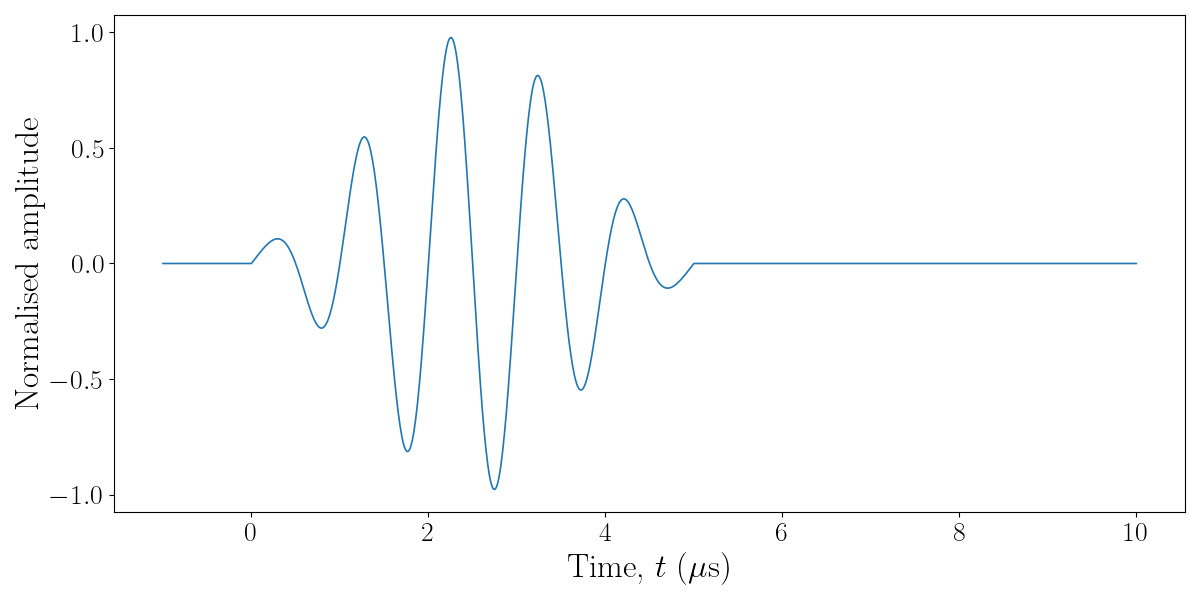}
	\caption{Driver signal applied to linear systems model of piezo-electric transducer actuator on the plate.}
	\label{fig:drive_sig}
\end{figure}

\begin{figure}[!ht]
	\centering
	\begin{subfigure}[h]{0.46\textwidth}
		\centering
		\includegraphics[width=\textwidth]{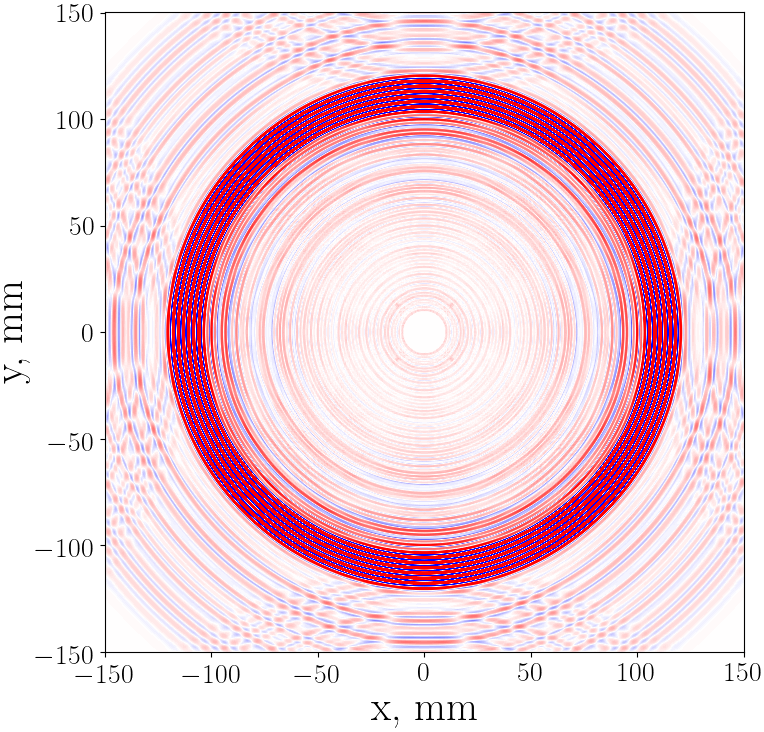}
		\caption{}
		\label{fig:frame_undam_35}
	\end{subfigure}
	\begin{subfigure}[h]{0.46\textwidth}
		\centering
		\includegraphics[width=\textwidth]{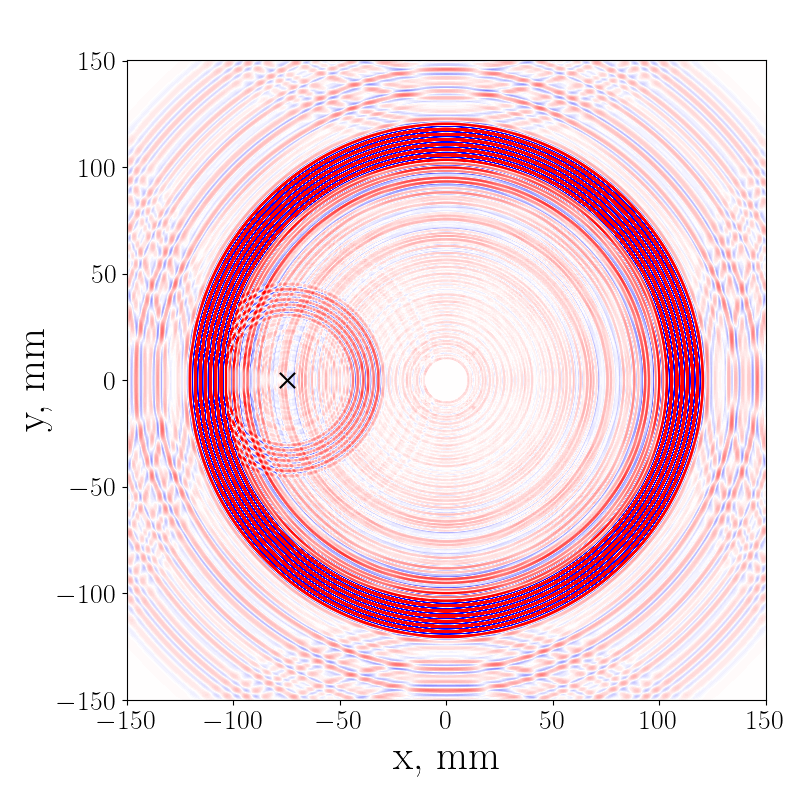}
		\caption{}
		\label{fig:frame_dam_35}
	\end{subfigure}
	\caption{Surface displacement results of LISA simulation for: (a) an undamaged plate and (b) a damaged plate (on which the location of damage is indicated by the black cross) $35\mu s$ from the beginning of the actuation signal. Dimensions are shifted to place the actuation source at the origin.}
	\label{fig:frames}
\end{figure}

Using the LISA approach here, Lamb waves were simulated in an aluminium plate, the details of which can be found in \Cref{tab:LISA_props}. %
First, a wave-field was simulated in an undamaged plate, then the simulation was repeated with a 1\si{mm} square, half-thickness notch at $x = 75$ \si{mm}, $y = 150$ \si{mm}. %
For all simulations, the actuation source is a 20mm diameter piezo-electric transducer (PZT) placed at the centre of the plate,  with a 1 \si{MHz}, five-cycle, Hanning-windowed sine wave. %
The driver signal applied to the linear systems model of the actuator is shown in \Cref{fig:drive_sig}. %
The actuation of the waves using a PZT is modelled using a linear systems transducer model, which approximates the pressure field output of the PZT when excited by a voltage \cite{Dobie2011}. %
Displacement of the points that are in contact with the PZT can then be calculated from this pressure field using the transmission coefficient between the plate and the PZT. %
For this work, only the layers of the plate and PZT are considered, and the adhesive layer between the two is ignored. %
The adhesive could be included relatively simply, though previous uses of this model show good correlation with experimental data \cite{Dobie2011}. %
The results of the damaged and undamaged plate simulations can be seen in \Cref{fig:frames}. %

\Cref{fig:frames} shows how the LISA simulation accurately models the effect of damage on the wave propagation, as the reflections of the $A_0$ mode can be clearly seen. %
It can be also be seen how the boundaries reflect the waves; this is an important consideration in the method, as this can have effects on the technique, as will be discussed later. %
The size of the damage here is considered relatively small, in comparison to the size of the plate, and so reflected waves will not be as strong as with other types of damage. %
This damage size was chosen to allow reflection that are relatively clear within the signal, but of which the amplitude is still relatively small compared to that of the nominal waves (see \Cref{fig:examp_dict}). %

\begin{figure}[h!]
	\centering
	\includegraphics[width=0.75\textwidth]{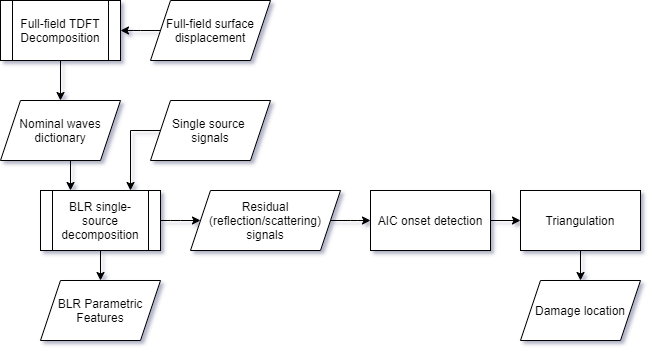}
	\caption{Localisation methodology.}
	\label{fig:method}
\end{figure}

\section{Methodology}
\label{sec:methodology}

The overall methodology for the localisation technique presented here is shown in the flowchart in \Cref{fig:method}. %
Before explaining the overall process, it useful to describe the data required for each stage of the methodology. %
For the baseline data -- in this case, forming the \emph{nominal wave dictionary}, which will be defined later -- time-space data of a wave along its propagation path is required. %
An example of such data is shown in \Cref{fig:time-distance}, and this comprises the ``Full-field surface displacement'' input in \Cref{fig:method}. %
For complex materials which have angle-dependent propagation characteristics would require data on multiple angles, though for the work in this paper an aluminium plate is considered for simplicity in the proof-of-concept study. %
The first stage of the process will produce a baseline dataset which is then used for the decomposition. %
The data required for the single-source decomposition stage is a time-signal at that location, and information on the sensor location in order to `align' the dictionary waves. %

The first key process comprises of using a \emph{full-field} decomposition, which uses the surface displacement data along the one-dimensional propagation path of the wave \cite{Haywood2020}. %
Data for this stage can be collected from either simulation or experimental regimes, the latter using a scanning laser doppler vibrometer (SLDV). %
An SLDV would be required for experimental acquisition of the full-field data, as the spatial resolution achievable with hard sensors, such as PZTs, would be insufficient. %
Though a small plate is used here, the method can be applied to any size plate, but with consideration of the spatial sampling size, and attenuation of the wave. %
The spatial sampling size is an important consideration in the experimental setup, as it directly relates to the wavenumber-bandwidth, as the reciprocal of the spatial sampling interval. %
Here, the second key process is the \emph{single-source} decomposition \cite{Haywood2021SPIE}, meaning decomposing a signal from one location, such as the voltage reading of a piezoelectric transducer. %
\Cref{sec:full-field-sep} begins here by explaining the details of the full-field multi-mode separation, which uses a forward-backward two-dimensional Fourier transform approach of masking the experimentally-determined dispersion curves. %
This stage produces a \emph{nominal wave dictionary} which provides the expected signal of individual wave modes for a given propagation distance. %
\Cref{sec:single-source-sep} details the single-source multi-mode separation method, which uses Bayesian linear regression (BLR) to attain a probabilistic distribution of the expected waves. %
Finally, \Cref{sec:localise_method} explains the triangulation procedure, and how the times of arrival of the reflected waves are determined using the Akaike Information Criterion (AIC) method. %

\subsection{Data sources}

\begin{figure}[h!]
	\centering
	\includegraphics[width=\textwidth]{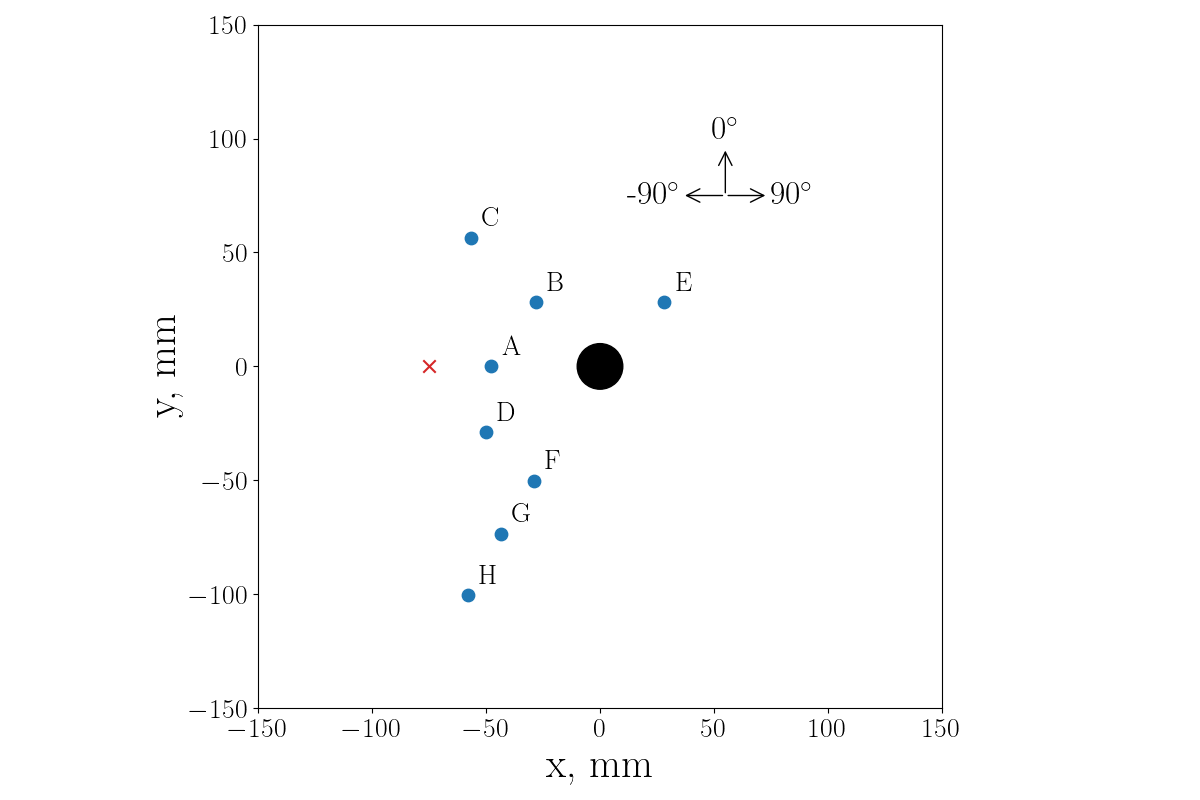}
	\caption{Sensor locations used for two-dimensional localisation; the black circle indicates the location of the PZT actuator, the blue circles indicate the locations of the `sensors' and their labels, and the red cross indicates the location of the damage. The arrows show the references for the direction of wave propagation from the actuator.}
	\label{fig:sens_locations}
\end{figure}

The aim of this subsection is to detail the source of the data used in this paper for each stage of the methodology. %
All data used here are taken from the LISA simulations described in \Cref{sec:LISA}. %
The simulation provides the full wave field information for all points on the surface of the plate, then data are extracted at each stage to represent the minimum data required for this stage. %
For the full-field multi-mode separation stage, surface displacement is taken from the undamaged plate simulation at regularly spaced intervals of 0.25 \si{mm} along the propagation path -90\si{\degree} from the origin (refer to \Cref{fig:sens_locations}). %
This subset gives full-field propagation data of the wave as it travels along one direction. %
For the single-source decomposition, the `sensor signals' are the surface displacements taken from the damaged simulation at single points. %
These are referred to as `measured signals' as they represent what would be recorded using PZT sensors in application. %
\Cref{fig:sens_locations} shows the locations of the sensor signals on the plate simulation outlined in \Cref{sec:LISA}. %
The sensor locations were chosen to give a variety of propagation and reflection angles, but not to be too close to the edge of the plates, as boundary reflections will interfere with the signal. %
For the purposes of sensor location description, the direction of the path between the actuation source and the damage is herein referred to as the actuator-damage direction, and the direction of the path between a sensor and the actuator is the incident direction of that sensor. %
The angle between these two directions is referred to as the reflection-incident angle (RIA). %
Sensor A has an RIA of 0\si{\degree}, sensors B and C have an RIA of 45\si{\degree}, sensor D has an RIA of 30\si{\degree}, sensor E has an RIA of 135\si\degree, and sensors F, G and H have an RIA of 60\si{\degree}. %
In order to localise a reflection source over a 2D plane, a minimum of three sensors is needed \cite{Kundu2014}, so during the localisation step various combinations of three sensors from the eight possible were chosen. %

The plate simulated in this paper is relatively small, and so there is a quick superposition of reflected waves within the signal at points further from the damage. %
In practice, the user must take careful consideration of reflective characteristics of the system, such as complex geometrical features, though this is a consistent trait of guided-wave based systems \cite{Kundu2014}. %

\subsection{Full-field multi-mode separation}
\label{sec:full-field-sep}

\begin{figure}[!ht]
	\centering
	\begin{subfigure}[h]{0.48\textwidth}
		\centering
		\includegraphics[width=\textwidth]{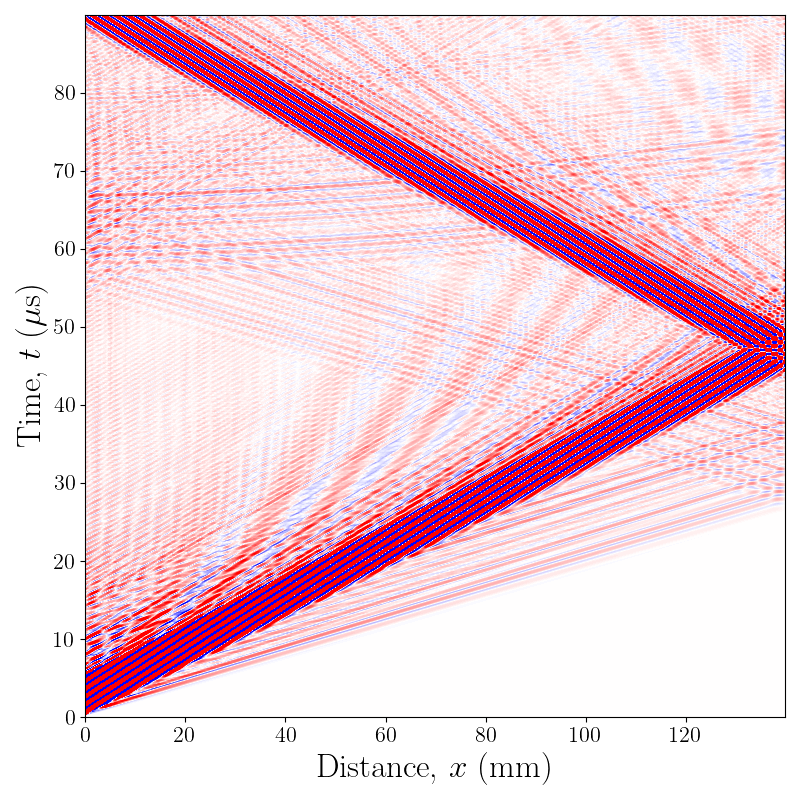}
		\caption{}
		\label{fig:time-distance}
	\end{subfigure}
	\begin{subfigure}[h]{0.48\textwidth}
		\centering
		\includegraphics[width=\textwidth]{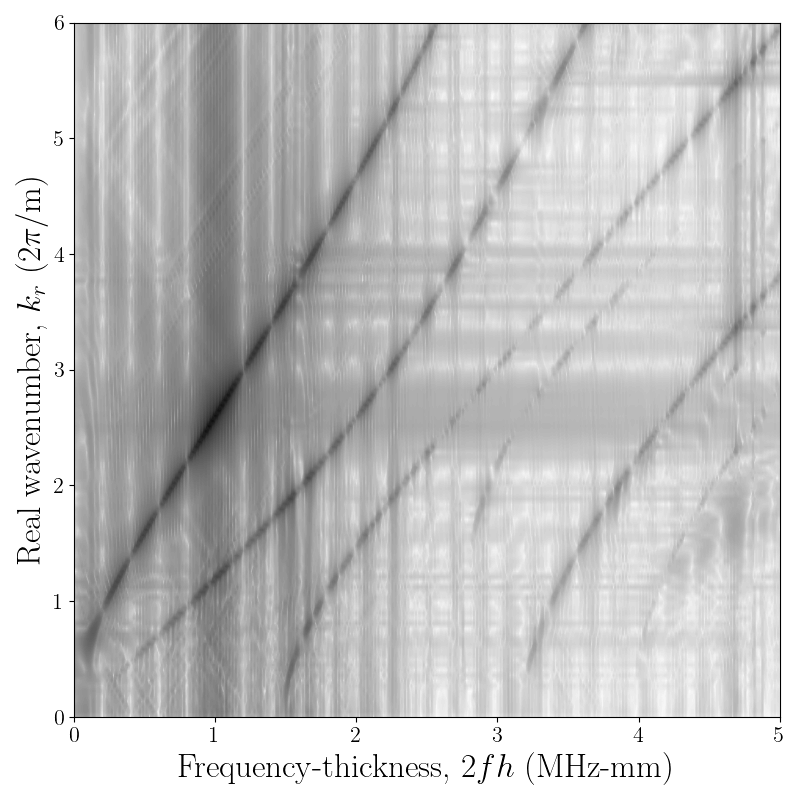}
		\caption{}
		\label{fig:frequency-wavenumber}
	\end{subfigure}
	\caption{Example of dispersion information determined from a TDFT by extracting an example of the [$\var{t-x}$] data from the LISA simulation, shown in terms of (a) $u(t,x)$ and (b) $P(f,k)$}
	\label{fig:TDFT_transform}
\end{figure}

Separation of multi-mode signals of a Lamb wave over its propagation in space is carried out by a forward-backward method using the experimentally-determined dispersion curve. %
The full-field decomposition is originally shown in \cite{Haywood2020}, as well as discussion of the advantages and caveats of the method. %
First the dispersion curve must be obtained by spatially sampling the surface displacement of a Lamb wave as it propagates through a bounded medium, then one performs a \emph{two-dimensional Fourier transform} (TDFT) along time and distance sequentially \cite{Alleyne1991}, 
\begin{equation}
	U(f, k) = \int_{-\infty}^{\infty} \int_{-\infty}^{\infty} u(t, x)\textrm{e}^{-\textrm{i}2\pi(ft+kx)} \;\textrm{dt}\;\textrm{dx}
	\label{eq:TDFT}
\end{equation}
where $u(t,x)$ is the surface displacement with respect to time $t$ and distance $x$, $f$ is the frequency in \si{Hz}, and $k$ is the wavenumber. %
In the discrete form this approach uses a \emph{two-dimensional fast Fourier transform} (TDFFT) algorithm,
\begin{equation}
	U[m, n] = \sum^{M-1}_{m=0} \sum^{N-1}_{n=0} u[m,n] \textrm{e}^{-\textrm{i}2\pi(fm/M+kn/N)}
	\label{eq:TDFFT}
\end{equation}
where $M$ and $N$ are the number of data points in time and distance respectively. %
From the simulation data described in \Cref{sec:LISA}, surface displacement data were extracted along a single propagation path. %
The spatially sampled signals are normalised, based on dividing each signal by the maximum amplitude of the signal, and then passed through the TDFFT algorithm to generate the [$\var{f-k}$] space. %
The resulting image data, which show dispersion curve information, are shown in \Cref{fig:TDFT_transform}. %
For the purposes of visualisation and ridge extraction, the [$\var{f-k}$] space is expressed in term of its magnitude, using a log transformation, 
\begin{equation}
	P(f,k) = 20\log_{10}(1+|U(f,k)|)
\end{equation}

Several distinct modes can be seen in the experimentally determined dispersion curve in \Cref{fig:frequency-wavenumber}. %
In order to separate the modes, the equivalent curves must first be extracted from the image data. %
This extraction was done here using a simple ridge-picking algorithm which extracts local maxima as points of the curve. %
The algorithm considered each pixel in the image and subtracted the mean of the surrounding $L^2$ pixels, it then normalised the data and set any point with a value below or equal to 50\% to 0, and any above 50\% to be 1. %
The resulting Boolean image data, $D$, are then used to extract the curves from the complex TDFFT data by simply performing an element-wise multiplication,
\begin{equation}
	U_* = D \circ U
	\label{eq:flatten}
\end{equation}
In order to reduce loss of information as much as possible from the [$\var{f-k}$] space, a buffer of $\pm a$ and $\pm b$ additional data points, in the horizontal and vertical directions respectively, are set to be 1 in $D$. %
The Boolean mask $D$ must be the same size as $U$, and it is important to select modes that appear on the full TDFFT data which represent the negative frequencies and wavenumbers in the transform; this can be done by simply returning the mirrored indices from the ridge selection process. %
$D$ can also be tuned to include any number of selected modes to determine which will be included in the reconstructed signal. %
Once these curves are extracted, they can be used to reconstruct individual, or selected, modes by applying an \emph{inverse two-dimensional Fourier transform} (ITDFT) on the new image data,
\begin{equation}
	u_*(t, x) = \int_{-\infty}^{\infty} \int_{-\infty}^{\infty} U_*(f, k)\textrm{e}^{\textrm{i}2\pi(ft+kx)} \;\textrm{df}\;\textrm{dk}
	\label{eq:ITDFT}
\end{equation}
or in discrete matrix form using an \emph{inverse two-dimensional fast Fourier transform} (ITDFFT),
\begin{equation}
	u_*[m,n] = \sum^{M-1}_{f=0} \sum^{N-1}_{k=0} U_*[f,k] \textrm{e}^{\textrm{i}2\pi(fm/M+kn/N)}
	\label{eq:ITDFFT}
\end{equation}
The modes that are included in the reconstructed signals were chosen based on comparison to numerical results of dispersion curves determined using DISPERSE \cite{Pavlakovic1997disperse}. %
The resulting decomposed waves were then stored as a \emph{nominal wave dictionary} (NWD) for use in the single-source decomposition stage. %
Although the signals are normalised at each distance, it is trivial to return the range of the decomposed signals to that of the measured signal by storing the normalisation parameters. %
For this work the signal data are kept at the normalised values as the decomposition method is focussed on just the \emph{shape} of the expected nominal signals and not their amplitude. %
The full-field multi-mode stage is summarised in \Cref{alg:full_field_sep}. %

\begin{algorithm}[h]
	\caption{\textsl{Full-field multi-mode separation.}}
	  \label{alg:full_field_sep}
	  \SetAlgoLined
	  \SetKwInOut{Input}{Input}\SetKwInOut{Output}{Output}
	  \Input{~~Multi-mode surface displacement signals over propagation distance $u$, initial guess of lowest $FTP$ and $k$ on dispersion curve for each mode $g$}
	  \Output{~~Individual nominal wave dictionary over propagation distance $u_*$}
	\BlankLine
	\BlankLine
	  \For{Distance point n}{
		  \textit{Normalise} signal $u(:,n) \rightarrow \hat{u}(:,n)$, storing peak-to-peak range $PTP(i)$\;
	  }
	  $U=TDFFT(\hat{u})$ using \cref{eq:TDFFT}\;
	  \textit{Power transform} of $U$: $P = 20\log_{10}(1+\textrm{abs}(U))$\;
	  \textit{Ridge finding}: get locations in $P$ where the normalised value is greater than 50\% above the mean of the surrounding $L^2$ pixels\;
	  \For{Wave mode g}{
	  \textit{Curve selection}: from initial guess of $\omega$ and $k$ for each mode, continuously select ridge points that are all less than $d_{\omega}$ and $d_{k}$ away in frequency and wavenumber respectively, set these locations in $D$ to equal 1\;
	  \textit{Flatten} dispersion curve image using \cref{eq:flatten}\;
	  \textit{Reconstruct} time-distance data for individual modes using \cref{eq:ITDFFT} and store in $u_*(g,:,:)$\;
	  }
	\label{alg:1}
\end{algorithm}

\subsection{Single-source multi-mode separation}
\label{sec:single-source-sep}

After the \emph{nominal wave dictionary} was determined and stored, it was used to decompose a Lamb wave signal from a single-source into the nominal and residual signals. %
This was done using a method first shown by Haywood-Alexander \textit{et al.}\ \cite{Haywood2021SPIE}. %
This method uses a Bayesian linear regression (BLR) technique, a description of which can be found in \cite{Murphy2012} and briefly in \cref{app:BLR}. %

For the purposes of this work, the model of the form,
\begin{equation}
    y(x,t) = \mathbf{w}(x)^{\top} \phi(x,t) + \varepsilon, \qquad \varepsilon \sim \mathcal{N} (0,\sigma^2)
    \label{eq:BLR_modelForm}
\end{equation}
uses basis form $\phi(x,t) = \{y_{A_0}(x,t),\; y_{S_0}(x,t)\}$, where $y_{\psi}(x,t)$ is the surface displacement of the normalised nominal wave for mode $\psi$ at propagation distance $x$ and time point $t$, $\mathbf{w}$ is the weight vector of the model, and $\varepsilon$ is the noise term normally distributed around 0 with a variance of $\sigma^2$. %
The nominal waves are also re-normalised to their individual ranges in order to better compare the predicted weights between the wave modes. %
One could consider $\phi$ to be the decomposition of the expected nominal signal into the selected modes. %
For this work, only the fundamental modes are used, although it would be trivial to use the same method on more selected modes. %
For this method, any signal range can be used, although it is preferable to use a normalised signal. %
The expected weight parameters $\mathbf{w}$ will give relative amplitudes of the nominal waves, directly influenced by the range of $y_i$. %
For computational reasons, it is preferable to use the normalised ranges for these wave vectors as any simultaneous loss in amplitude of the nominal waves will not be captured. %

\begin{figure}[h]
	\centering
	\includegraphics[width=0.9\textwidth]{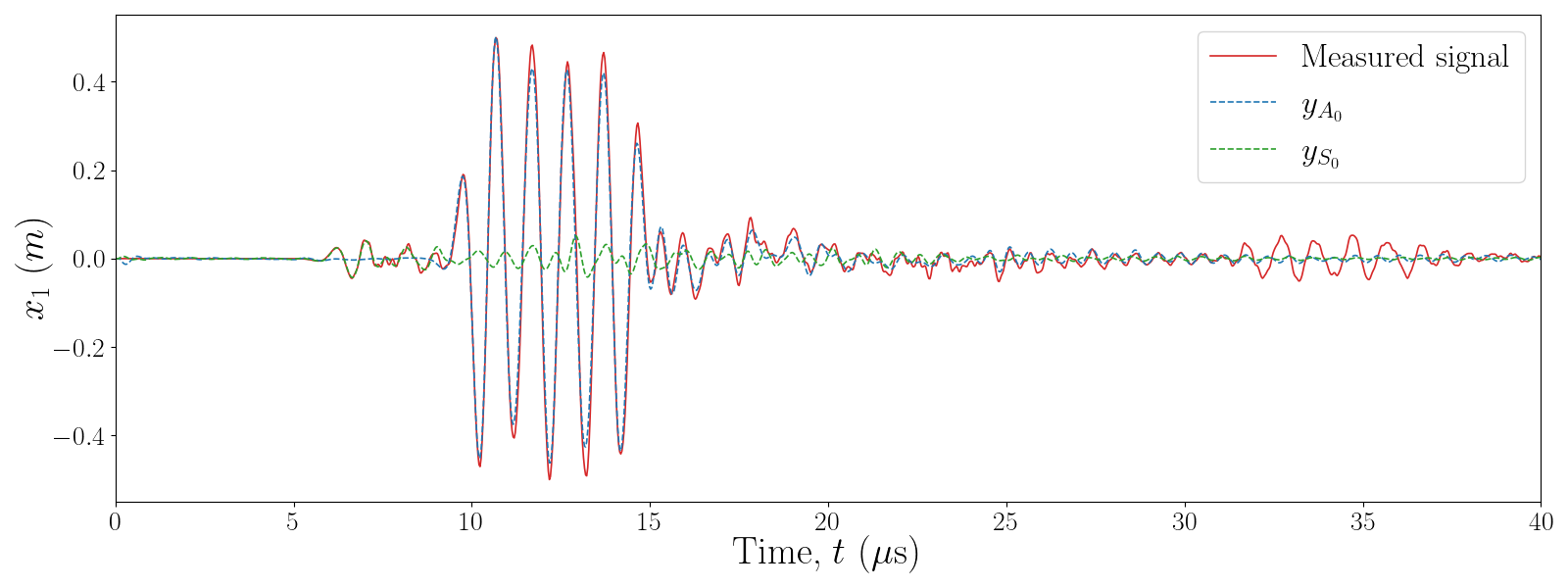}
	\caption{Example nominal wave dictionary signals $y_{A_0}$ and $y_{S_0}$ at a propagation distance $x$ of 50 \si{mm}, compared to the measured signal in the damaged simulation. The dictionary signals are in their normalised values directly from the full-field decomposition and the measured signal is normalised using \cref{eq:normalise}.}
	\label{fig:examp_dict}
\end{figure}

For numerical reasons during the BLR decomposition, the signal is first normalised based on the range of the signal,
\begin{equation}
	\mathbf{y}^* = \frac{\mathbf{y}}{\text{range}(\mathbf{y})}
	\label{eq:normalise}
\end{equation}
where $\text{range}(\mathbf{y}) = \text{max}(\mathbf{y}) - \text{min}(\mathbf{y})$ is the peak-to-peak range of measured signal $\mathbf{y}$. %
The value of $\text{range}(\mathbf{y})$ is stored and the calculated weights are multiplied by this range to present them in terms of the signal amplitude in \si{m}. %
An example of the nominal wave dictionary signals used to decompose a measured signal on the damaged plate is shown in \Cref{fig:examp_dict}. %

An important metric that is attainable from the Bayesian linear regression method is the \emph{predictive likelihood} (PLL), which gives an indication of the likelihood that the model fits and takes into account the uncertainty as well as the quality of the mean fit. %
The definition and how this metric is calculated is given in \cref{app:BLR}. %
The single-source decomposition process is summarised in \Cref{alg:sing_source_sep}. %

\begin{algorithm}[h]
	\caption{\textsl{Single-source multi-mode separation.}}
	  \label{alg:sing_source_sep}
	  \SetAlgoLined
	  \SetKwInOut{Input}{Input}\SetKwInOut{Output}{Output}
	  \Input{~~Individual nominal wave dictionary over propagation distance $u_*$, measured signal $\mathbf{y}$ at propagation distance $\hat{x}$}
	  \Output{~~Predicted nominal wave signal at measured signal location $\tilde{\mathbf{y}}$, predicted variance of nominal wave signal $\sigma^2$}
	\BlankLine
	\BlankLine
	  \textit{Normalise} measured signal $\mathbf{y} \rightarrow \mathbf{y}^*$, storing \emph{peak-to-peak range} $PTP$\;
	  \textit{Load} wave mode signals for distance point $\hat{x}$ from nominal wave dictionary $u_*(:,:,x_*)$ based on $x_* = \text{argmin}(x - \hat{x})$\;
	  \textit{Sum} across modes to form the basis expansion $\mathbf{X} = \phi(x,:)$\;
	  \textit{Calculate} analytical solution of weights $\mathbf{w}$ and predicted variance $\sigma^2 = \textrm{diag}(\mathbb{V})$ (refer to \cref{alg:BLR})\;
	  \textit{Reconstruct} predicted nominal signal $\tilde{\mathbf{y}} = \mathbf{w}^{\top}\mathbf{X}$
	\label{alg:2}
\end{algorithm}

\subsection{Reflection source triangulation}
\label{sec:localise_method}

Once the signals are decomposed into their individual nominal modes, the residual signal can be calculated,
\begin{equation}
	\mathbf{y}_{res} = \mathbf{y} - \tilde{\mathbf{y}}
	\label{eq:residual}
\end{equation}
where $\mathbf{y}$ is the measured signal and $\tilde{\mathbf{y}}$ is the predicted nominal signal. %
As shown in \cite{Haywood2021SPIE}, this residual contains information on reflected/scattered waves because of damage. %
In order to triangulate the source of this reflection, the time of arrival at each sensor must first be determined. %
Previous work \cite{Holford2017} has shown a simple, yet effective technique using the Akaike Information Criterion (AIC) \cite{akaike1998markovian} to determine arrival times. %
The AIC function compares signal entropy before and after each time $t$ in a time series. %
When time $t$ is aligned with the signal arrival, the similarity between the high-entropy uncorrelated noise prior to $t$ and the low-entropy structured signal after $t$ is at its lowest and the function returns a minimum. %
Therefore, a simple minimum finding function can then be used to determine the location of the onset of a signal. %
The AIC of a signal $x$ for a given time point $t$ is given by,
\begin{equation}
	AIC(t) = t\log_{10}(\mathbb{V}\left[x_{1:t}\right]) + (T-t-1)\log_{10}(\mathbb{V}\left[x_{t:T}\right])
	\label{eq:AIC}
\end{equation}
where $\mathbb{V}\left[x_{j:k}\right]$ is the classic variance of $x$ from point $j$ to point $k$ and $T$ is the final time point in the series. %
For detection of reflection onset only, the portion of the signal after arrival of the slowest nominal wave was used. %
This selection is because onset detection by AIC is based on a change in entropy ratio in the signal before and after the onset, and there is increased structure in the residual at the nominal wave portions \cite{Haywood2021SPIE}. %

This onset then gives a reflection time of arrival at each sensor $t_{ref}$. %
It is useful to look at the problem of localising with a sensor along the propagation path between the actuation source and the damage. %
For this one-dimensional localisation problem, the distance of the reflection source from the sensor can be calculated simply as,
\begin{equation}
	d_r = \frac{1}{2}\frac{t_{ref}-t_{A0}}{c_{A0}}
	\label{eq:1d_local}
\end{equation}
where $t_{A0}$ is the time of arrival of the nominal $A_0$ mode at the sensor and $c_{A0}$ is the group velocity of the $A_0$ mode. %
It should be noted that this may not be a useful method in practice, as it is unknown whether the sensor, actuator and damage are collinear. %
Furthermore, \Cref{eq:1d_local} is only applicable when the sensor is earlier in the propagation path, because in locations after damage there are no additional waves, but only converted/reduced energy modes, as can be seen in \Cref{fig:frame_dam_35}. %
The one-dimensional procedure is included to show a simple illustrative example of localisation, as well as providing an initial stage to assess the decomposition-based approach to localisation. %
This stage is also included to allow the reader to gain intuition on the procedure and how a usable algorithm could be developed on this basis; a two-dimensional example of which will be demonstrated in this paper. %

As the Lamb wave is propagating in an isotropic, homogeneous structure, the difference in time of arrival between the sensors and wave velocity $c$ (prior knowledge), can be used to triangulate the source of the wave \cite{Tobias1976, Kundu2014}. %
For three sensors, first the reflection time of arrival for sensors 1, 2 and 3 are denoted as $t_{r,1}$, $t_{r,2}$ and $t_{r,3}$ respectively. %
The exact time of the reflection from the damage $t_0$ is unknown, and therefore so is the time taken for the reflected wave to travel from the damage to the sensors. %
These unknown times are denoted $t_1$, $t_2$, $t_3$ and the distance from the damage to each sensor is given by,
\begin{equation}
	d_i = c \times t_i
	\label{eq:di}
\end{equation}
As $t_i$ is unknown, $d_i$ cannot be calculated from \Cref{eq:di}. However, the difference in time of arrival (dTOA) between two sensors $t_{ij}$ provides the difference in distance between the damage location and sensors $i$ and $j$,
\begin{equation}
	d_{ij} = c \times t_{ij} = c \times (t_i - t_j)
\end{equation}

In this work, the triangulation of the reflection source is framed as an optimisation problem, as shown in \cite{Miller2005}. %
In this sense, the source location is estimated by minimising the difference between the recorded dTOA and a calculated value originating from a trial source position. %
In practice, this was done by defining a cost function,
\begin{equation}
	J = \sum_{i, j} \left(t_{ij} - \frac{|E - S_i|-|E-S_j|}{c}\right)^2
	\label{eq:local_opt}
\end{equation}
where $E$ is a trial location of the reflected signal source and $|E-S_n|$ is the Euclidean distance between this trial location and the location of sensor $n$. %
An estimate of the location of the reflection source was then determined by,
\begin{equation}
	\hat{E} = \textrm{arg min}(J)
\end{equation}
In this work, this minimisation uses the Nelder-Mead method \cite{Olsson1975}. %
The reflection triangulation process using various sensor locations is summarised in \Cref{alg:triangulation}. %

\begin{algorithm}[ht]
	\caption{\textsl{Reflection source triangulation.}}
	  \label{alg:triangulation}
	  \SetAlgoLined
	  \SetKwInOut{Input}{Input}\SetKwInOut{Output}{Output}
	  \Input{~~Measured signal at various locations $\mathbf{y}$, nominal wave dictionary over propagation distance $u_*$}
	  \Output{~~Predicted location of reflection source $\hat{E}$}
	\BlankLine
	\BlankLine
	  \For{Measured signal of sensor $i$}{
		  \textit{Decompose} signal using \cref{alg:sing_source_sep} to retrieve predicted nominal wave signal $\tilde{\mathbf{y}}$\;
			\textit{Construct} residual signal using \cref{eq:residual}\;
			\textit{Determine} onset of reflection using AIC $\rightarrow t_i$\;
	  }
	  \textit{Minimise} \cref{eq:local_opt} using \emph{Nelder-Mead} method to determine predicted location $\hat{E}$\;
	\label{alg:3}
\end{algorithm}

The algorithms are shown individually here for each stage but in practice work concurrently and data is fed from through each in order. %
Firstly, the full-field wave propagation data is fed into \Cref{alg:1}, from which the nominal wave dictionary $u_*$ is returned. %
This dictionary is then fed into \Cref{alg:2}, along with the single-source signals from each sensor $\mathbf{y}_i,\; i=1,2,3$, and the predicted nominal signals $\tilde{\mathbf{y}}_i$ are returned. %
The predicted nominal signals are then used in \Cref{alg:3} to return the predicted location of the damage, $\hat{E}$. %


\section{Probabilistic Decomposition Results and Discussion}
\label{sec:decomp_results}

To begin, it is useful to analyse and evaluate the probabilistic decomposition of the full-field propagation and single-source signals. %
This section initially looks at the decomposition of the full-field signal as shown in \Cref{sec:ff_decomp_results}, and then at various distances along the propagation path collinear with damage, the results of which are shown in \Cref{sec:ss_decomp_results}. %
Decomposition results at other propagation angles are then explored at the same distance, all from the damaged plate, in \Cref{tab:ss_decomp_metrics}. %

\subsection{Full-field decomposition}
\label{sec:ff_decomp_results}
\begin{figure}[h!]
	\centering
	\begin{subfigure}{0.48\textwidth}
		\centering
		\includegraphics[width=\textwidth]{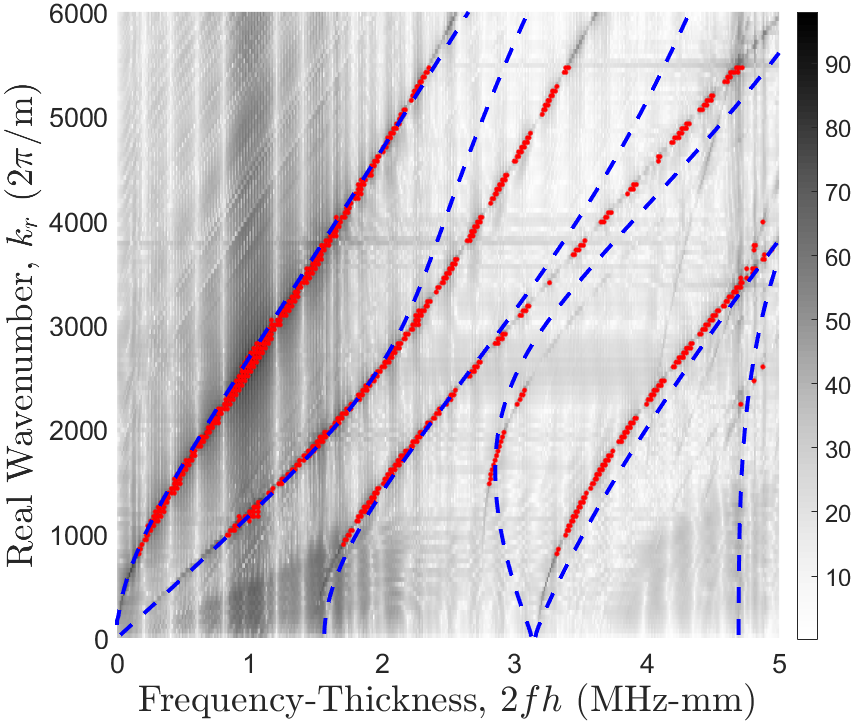}
		\caption{}
		\label{fig:ridgeSelections}
	\end{subfigure}
	\begin{subfigure}{0.48\textwidth}
		\centering
		\includegraphics[width=\textwidth]{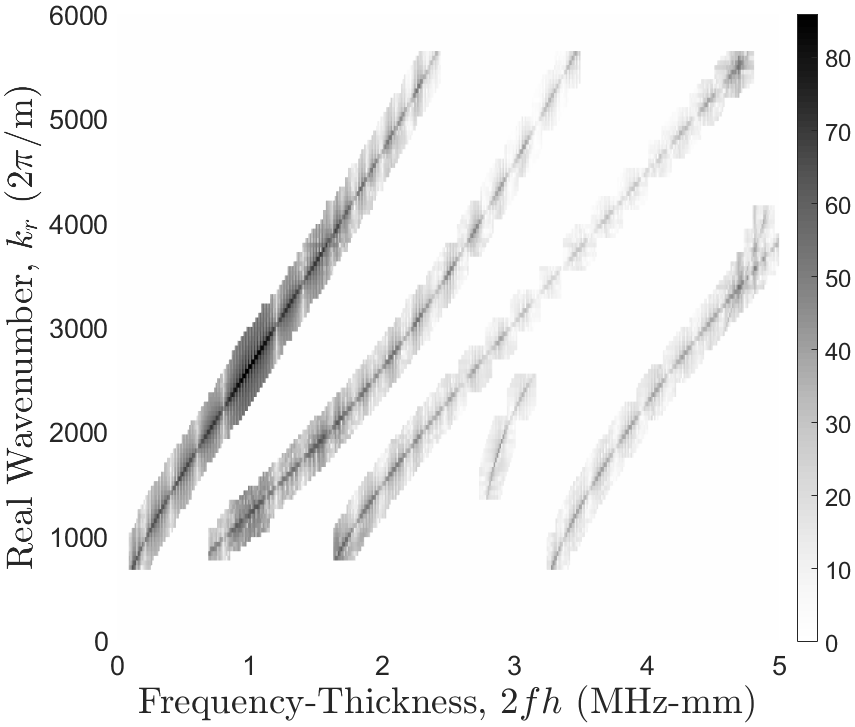}
		\caption{}
		\label{fig:flatDisp}
	\end{subfigure}
	\caption{(a) Point data of individual modes found using ridge-picking algorithm (red circle markers) and comparing to data from `Disperse' (blue dashed line). These lines are representative of (in order of appearance with increasing $2fh$) the $A_0,\;S_0,\;A_1,\;S_1,\;S_2$ and $A_2$ modes. (b) Separated dispersion curves extracted using a buffer of $a=11$ and $b=3$.}
\end{figure}

The results of passing the data shown in \Cref{fig:frequency-wavenumber} through the ridge picking algorithm are shown in \Cref{fig:ridgeSelections}. %
By comparing the dispersion curves generated using \Cref{eq:TDFT} to those generated with `DISPERSE', it can be seen which individual wave modes correspond to each ridge. %
The dispersion curves seen in the image data taken from the TDFFT show mostly good correspondence with those calculated numerically, discrepancies between them are due to the cells per unit wavelength \cite{Dobie2011}. %

After comparing the extracted ridges with the numerically-determined dispersion curve data to label the curves selected, the image data was flattened for each mode using \Cref{eq:flatten}. %
The separated mode data, $H_*$, are shown in \Cref{fig:flatDisp} in terms of $P(f,k)$, with the $A_0$, $S_0$, $A_1$, $S_1$, $S_2$ modes selected. %

\begin{figure}[h!]
	\centering
	\includegraphics[width=\textwidth]{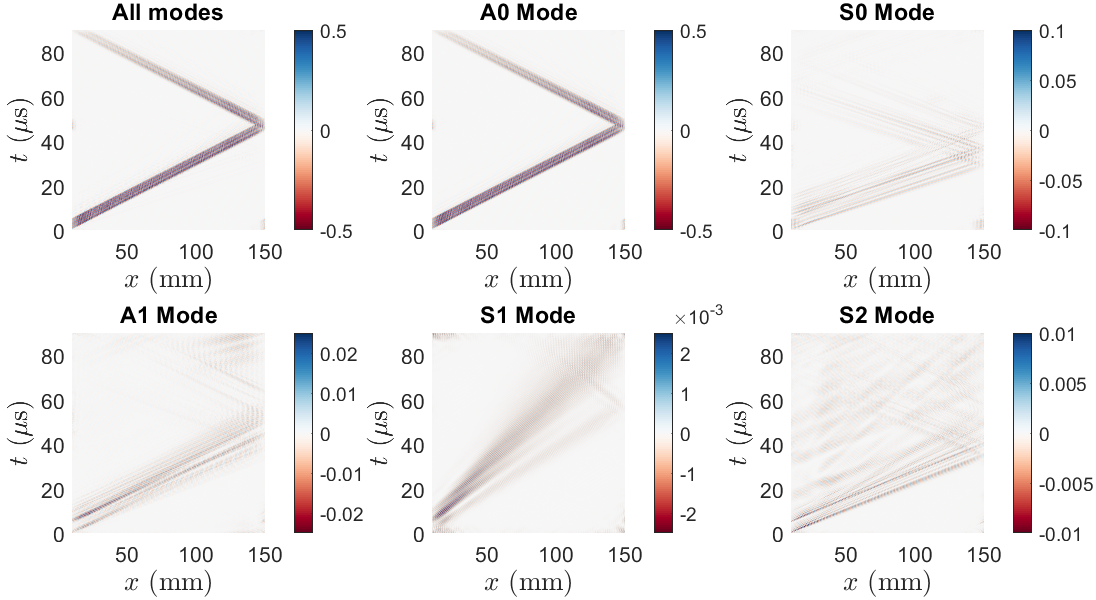}
	\caption{Reconstructed normalised time-distance spaces using inverse TDFFT for all modes together and each mode individually.}
	\label{fig:txReconstr}
\end{figure}

\begin{figure}[h!]
	\centering
	\includegraphics[width=0.8\textwidth]{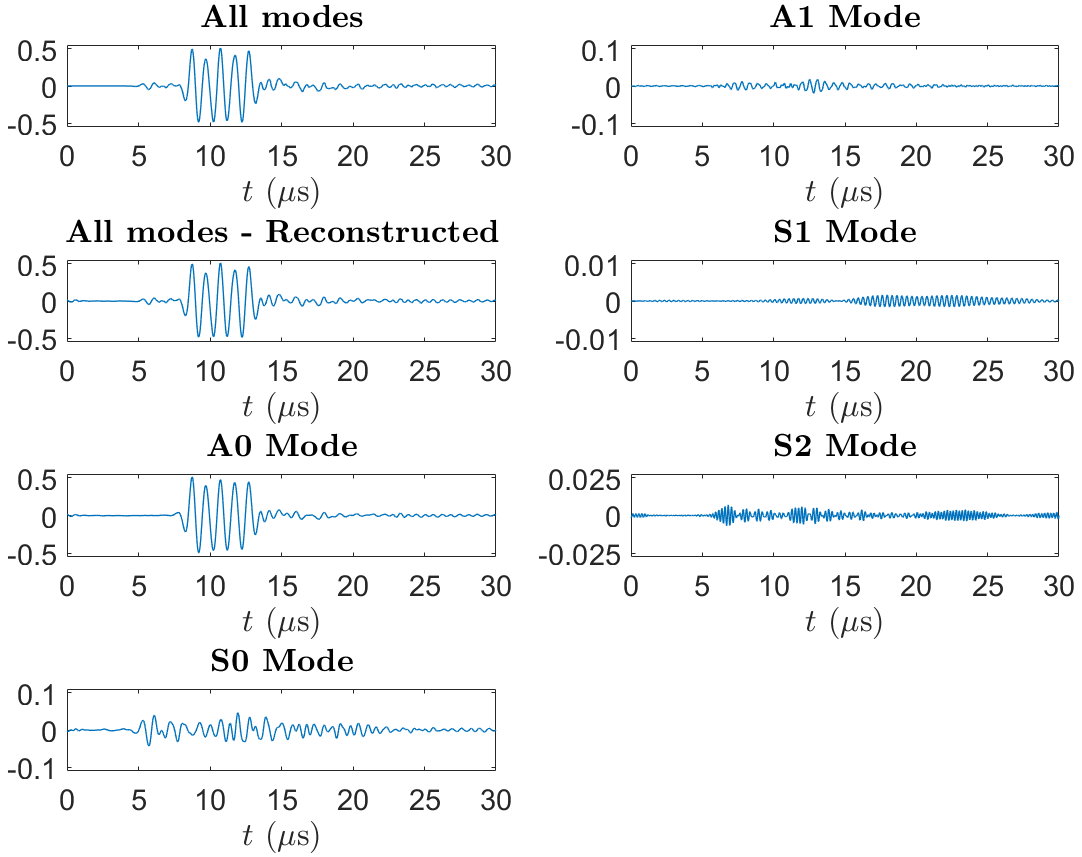}
	\caption{Individual normalised signals taken from propagation distance of 50mm for original data and reconstructed signals for all modes together and each mode individually.}
	\label{fig:reconstSignals}
\end{figure}

The full-field wave dictionary $u_*$ was then generated by passing each flattened mode through \Cref{eq:ITDFT} in series. %
The total reconstructed time-distance space can be seen in \Cref{fig:txReconstr}, and individual time signals at a propagation distance of 10mm in \Cref{fig:reconstSignals}. %
In comparison to the original data in \Cref{fig:time-distance}, the reconstructed [$\var{t-x}$] space is missing a larger wavelength component which is likely to be a mode that is not visible in \Cref{fig:ridgeSelections}. %
From \Cref{fig:reconstSignals} it can be seen that the relative amplitudes of the individual wave modes is preserved throughout the separation. %
As the amplitudes of the $A_0$ and $S_0$ modes are significantly larger than the higher order modes, it is reasonable to only take these two modes forward to be used in the separation. %
Thus the nominal wave dictionary $u_*$ is reduced to $u_*\{A_0,S_0\}$. %

\Cref{fig:reconstSignals} also shows that the $A_0$ mode is the most influenced by the actuation signal and is the signal which retains the shape of this the most. %
This observation can be explained by examining the dispersion curves; the actuation source was driven at 300kHz-mm, at which there are only two solutions for the wavenumber, the $A_0$ and $S_0$ modes. %
Therefore, the actuation of higher-order modes is due to excitation of wavelength components from the fundamental modes. %

In \Cref{fig:reconstSignals}, one can see the original and reconstructed signals show the multi-spectra behaviour of the influence of multi-modal signals in individual wave modes. %
As the plate is excited with a single PZT on the upper surface, this primarily actuates the antisymmetric mode, shown as this wave shape is most similar to that of the actuation \Cref{fig:drive_sig}. %
Without full-field multi-mode simulation, it has not been shown how to model this inidividual signal analytically. %

\subsection{Single-source signal decomposition}
\label{sec:ss_decomp_results}

\begin{figure}[h!]
	\centering
	\begin{subfigure}{0.49\textwidth}
		\includegraphics[width=\textwidth]{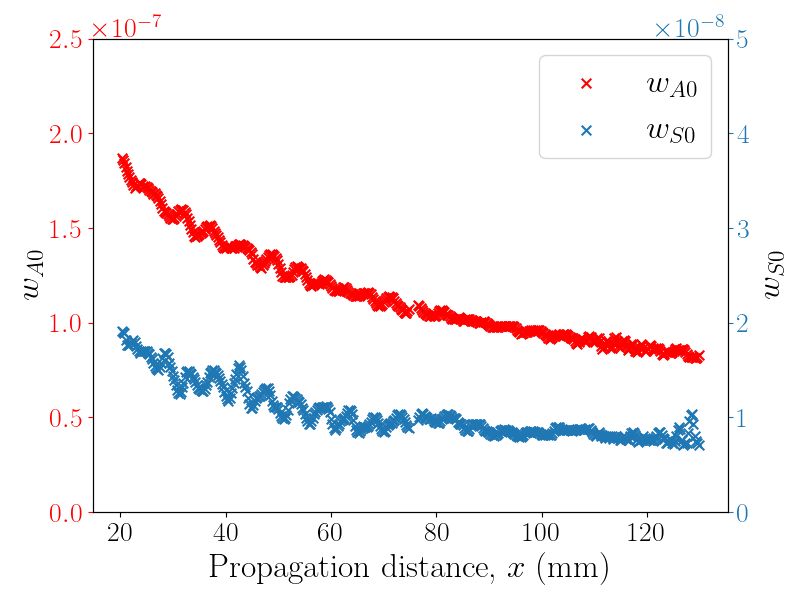}
		\caption{}
		\label{fig:BLR_weights_undam}
	\end{subfigure}
	\begin{subfigure}{0.49\textwidth}
		\includegraphics[width=\textwidth]{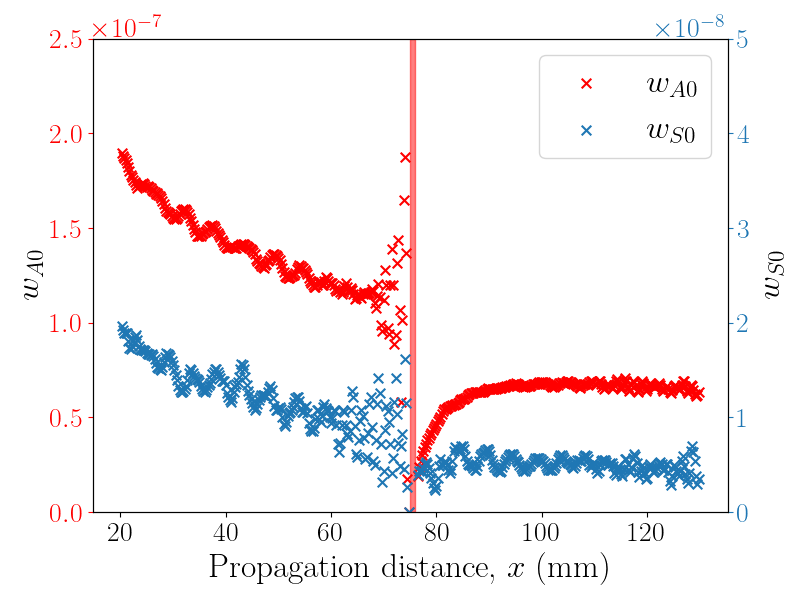}
		\caption{}
		\label{fig:BLR_weights_dam}
	\end{subfigure}
	\caption{Weights representing each mode extracted from BLR decomposition applied at various propagation distances for (a) the undamaged plate and (b) damaged plate. In (b) the section of propagation in which there is damage is given by the vertical filled red area.}
	\label{fig:BLR_weights}
\end{figure}


The calculated weights from the BLR decomposition are shown in \Cref{fig:BLR_weights}, where the weights are written as the corresponding weight for each mode, $w_{A0}$ and $w_{S0}$. %
The propagation location in which there is damage is given by the filled red area in \Cref{fig:BLR_weights_dam}. %
These weights can be interpreted as parameters relating the shape of the expected nominal wave to the range of the measured nominal wave. %
It is clear to see the attenuation of each mode as the propagation distance increases, although an interesting point to note is the sharp drop in the antisymmetric parameter $w_{A0}$, after the wave has propagated through the damage. %
There is a periodic nature to the weights with respect to the propagation distance; this is likely to be because of reflected waves coming in and out of phase with the nominal waves. %
The superposition of these waves will affect the amplitude of the nominal waves as the phase differences between the waves change. %

\begin{figure}[h!]
	\centering
	\begin{subfigure}{0.49\textwidth}
		\includegraphics[width=\textwidth]{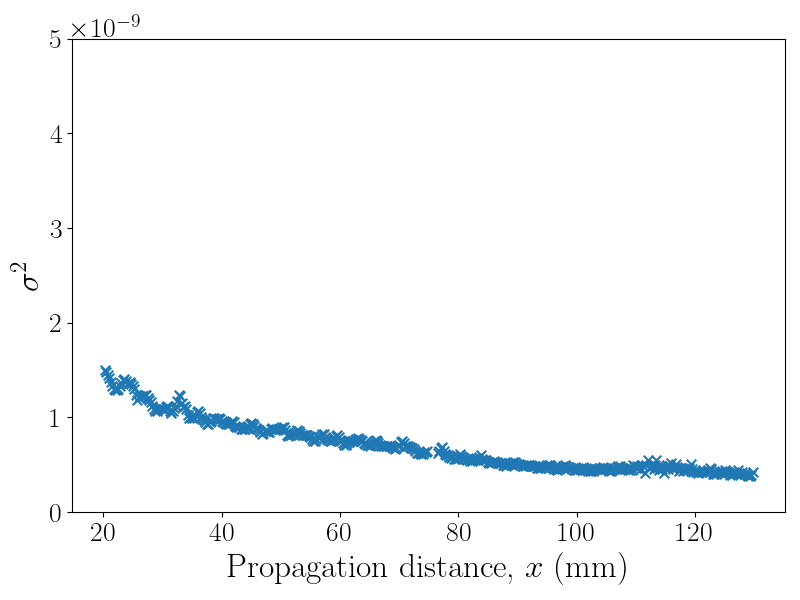}
		\caption{}
		\label{fig:BLR_sigma_undam}
	\end{subfigure}
	\begin{subfigure}{0.49\textwidth}
		\includegraphics[width=\textwidth]{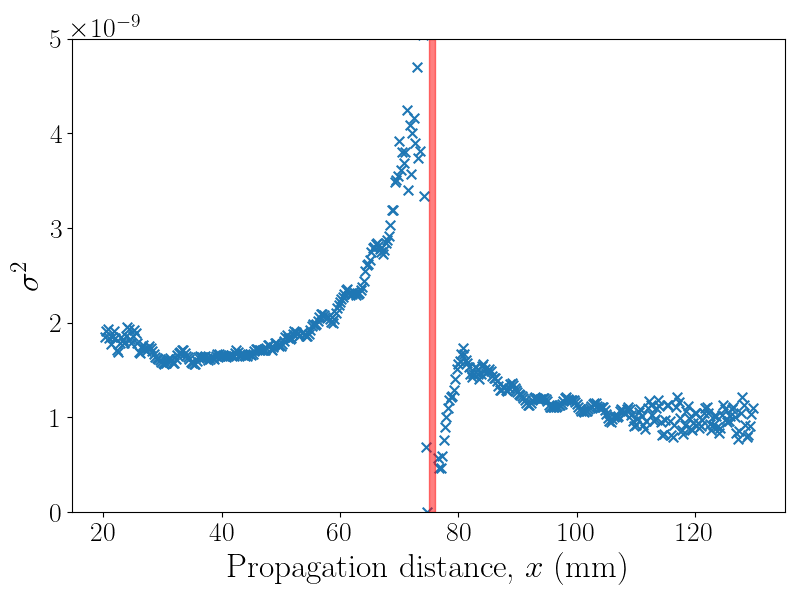}
		\caption{}
		\label{fig:BLR_sigma_dam}
	\end{subfigure}
	\caption{Predicted variance of nominal waves extracted from BLR decomposition applied at various propagation distances for (a) the undamaged plate and (b) damaged plate. In (b) the section of propagation in which there is damage is given by the vertical filled area.}
	\label{fig:BLR_sigma}
\end{figure}

\Cref{fig:BLR_sigma} shows the predicted variance of the decomposed signals at each propagation distance. %
An initial observation is the increased predicted variance when decomposing signals in the plate containing damage. %
Furthermore, in \Cref{fig:BLR_sigma_dam} the predicted variance appears to increase as one moves closer to the reflection source, and there is a sharp drop afterwards. %
This trend can be explained using the information viewed in \Cref{fig:BLR_weights_dam}; as the amplitude of the nominal wave decreases closer to the damage, and any reflected waves will also attenuate further away from damage, the relative influence of the superposition of reflected waves will increase closer to damage. %
However, past the damage there is no reflection in one plane, instead there is mode conversion. %
Therefore, the predicted variance will drop along with the magnitude of the predicted weights as seen in \Cref{fig:BLR_weights_undam}. %


\begin{table}[h!]
	\centering
	\begin{tabular}{| c | c | c | c | c | c | c |}
		\hline
		Angle & Distance & $w_{A0}$ & $w_{S0}$ & $\sigma^2$ & $PLL$ \\
		\hline
		-90\si{\degree} & 50 \si{mm} & 1.397e-07 & 1.337e-08 & 1.645e-09 & -1.946e+09 \\
		90\si{\degree} 	& 50 \si{mm} & 1.396e-07 & 1.228e-08 & 9.624e-10 & -4.207e+08 \\
		-45\si{\degree} & 50 \si{mm} & 1.157e-07 & 1.073e-08 & 3.224e-09 & -2.152e+09 \\
		-60\si{\degree} & 50 \si{mm} & 1.184e-07 & 5.359e-09 & 2.773e-09 & -6.516e+08 \\
		\hline
	\end{tabular}
	\caption{Resulting metrics from the BLR decomposition of single-source signals at various locations over the damaged plate, noted by their propagation angle and distance. A propagation angle of $\textrm{-}90^{\circ}$ is the direction of damage from the actuation source. Here, $w_{A0}$ and $w_{S0}$ represent the weight parameters to construct the nominal signals of the $A_0$ and $S_0$ mode respectively. }
	\label{tab:ss_decomp_metrics}
\end{table}

\begin{figure}[ht]
	\centering
	\includegraphics[width=0.9\textwidth]{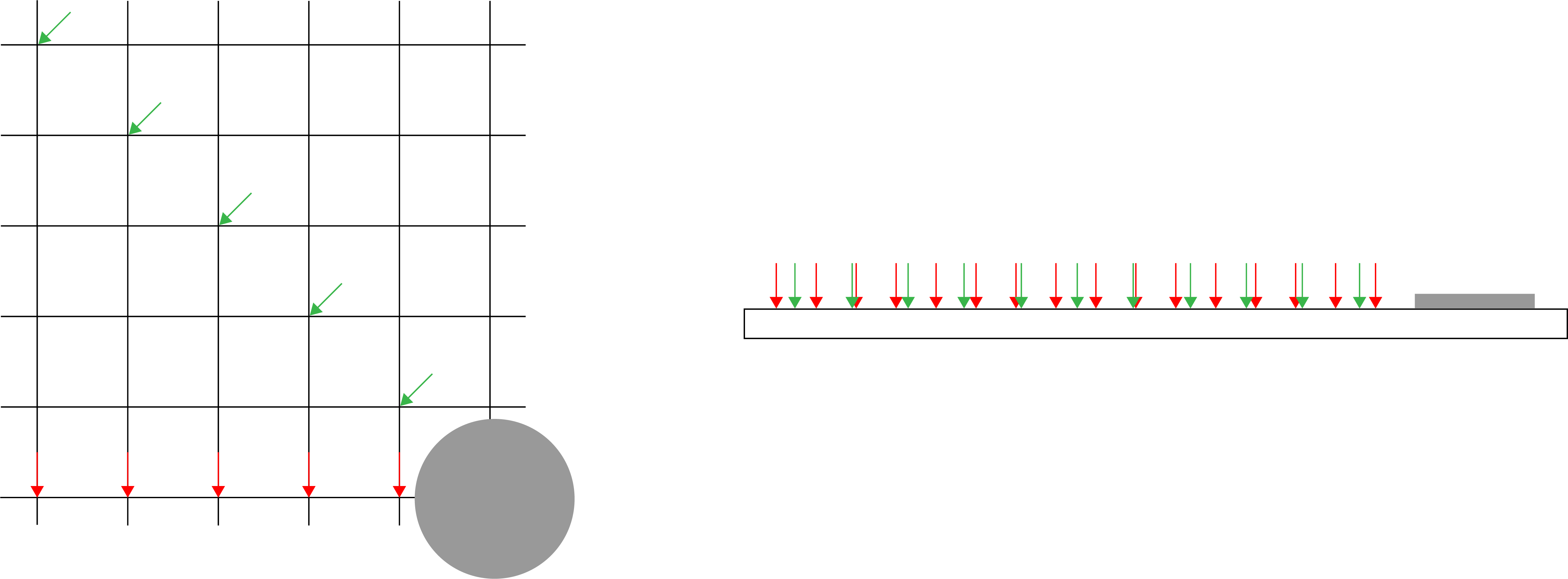}
	\caption{Illustration of how propagation angle influences spatial sampling of surface displacement along the propagation path. The grey circle represents the actuator source (not to scale). The left figure shows, with red arrows, the location in the LISA simulation of consecutive points at -90\si{\degree} from the actuation source and in green arrows consecutive points at -45\si{\degree}. The right figure compares the sampling pattern of the consecutive points at different angles.}
	\label{fig:spatial_sampling}
\end{figure}

Surface displacement signals were taken from the damaged plate at various propagation angles and the decomposition technique applied to each of these. %
The resulting metrics returned by the technique are shown for each signal location in \Cref{tab:ss_decomp_metrics}. %
A notable observation from these results is the increased uncertainty when decomposing signals not taken along the same propagation angle at which the full-field decomposition took place. %
This increased uncertainty is likely because of the change in spatial sampling; the data used for full-field decomposition was taken along the -90\si{\degree} propagation path which will result in the smallest spatial sampling step size equal to the cell dimension $\varepsilon$.
An illustration of how the selected propagation angle influences spatial sampling is shown in \Cref{fig:spatial_sampling}, where it can be seen that, when data are taken at a propagation angle of -45\si{\degree}, spatial sampling step size is at a maximum. %

\begin{figure}[ht]
	\centering
	\includegraphics[width=0.9\textwidth]{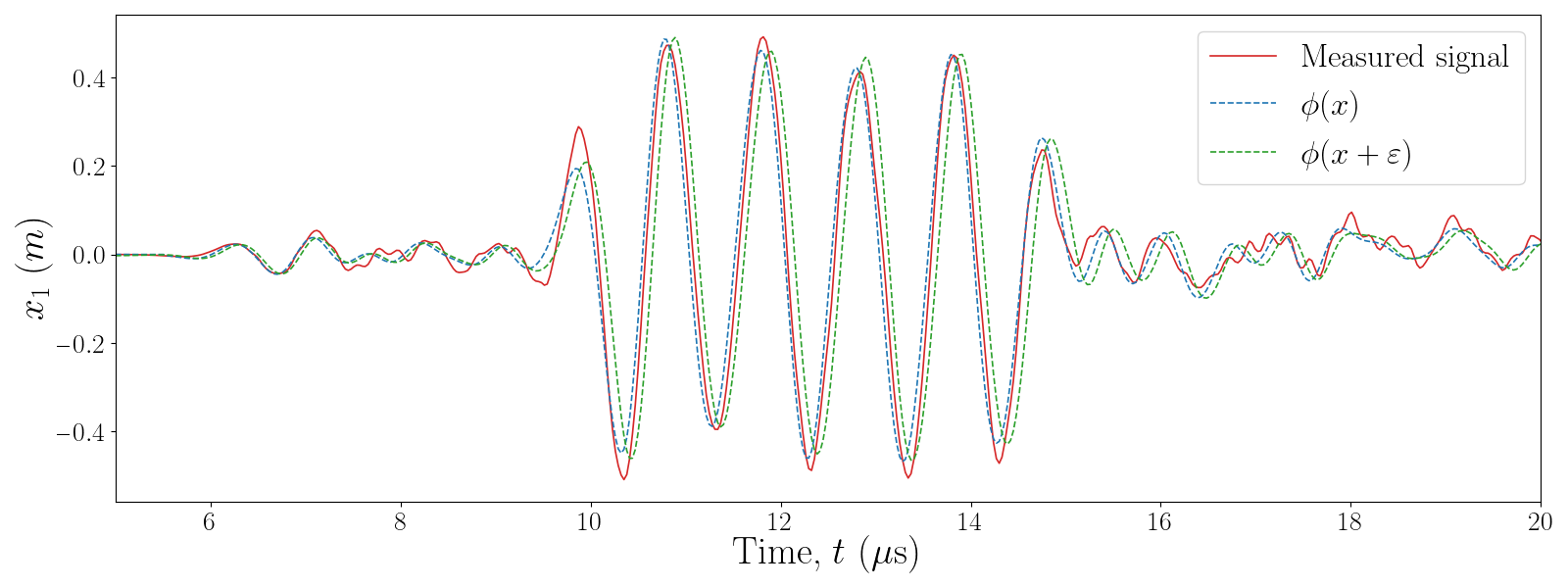}
	\caption{Comparison of the measured signal at a propagation distance and angle of 50 \si{mm} and 45\si{\degree} respectively, and the dictionary wave $\phi(x)$ used for decomposition at this point. In addition, the dictionary wave at one distance point further $\phi(x+\varepsilon)$ is shown.}
	\label{fig:dict_align_phase}
\end{figure}

\begin{figure}[h!]
	\centering
	\begin{subfigure}[h]{0.95\textwidth}
		\includegraphics[width=\textwidth]{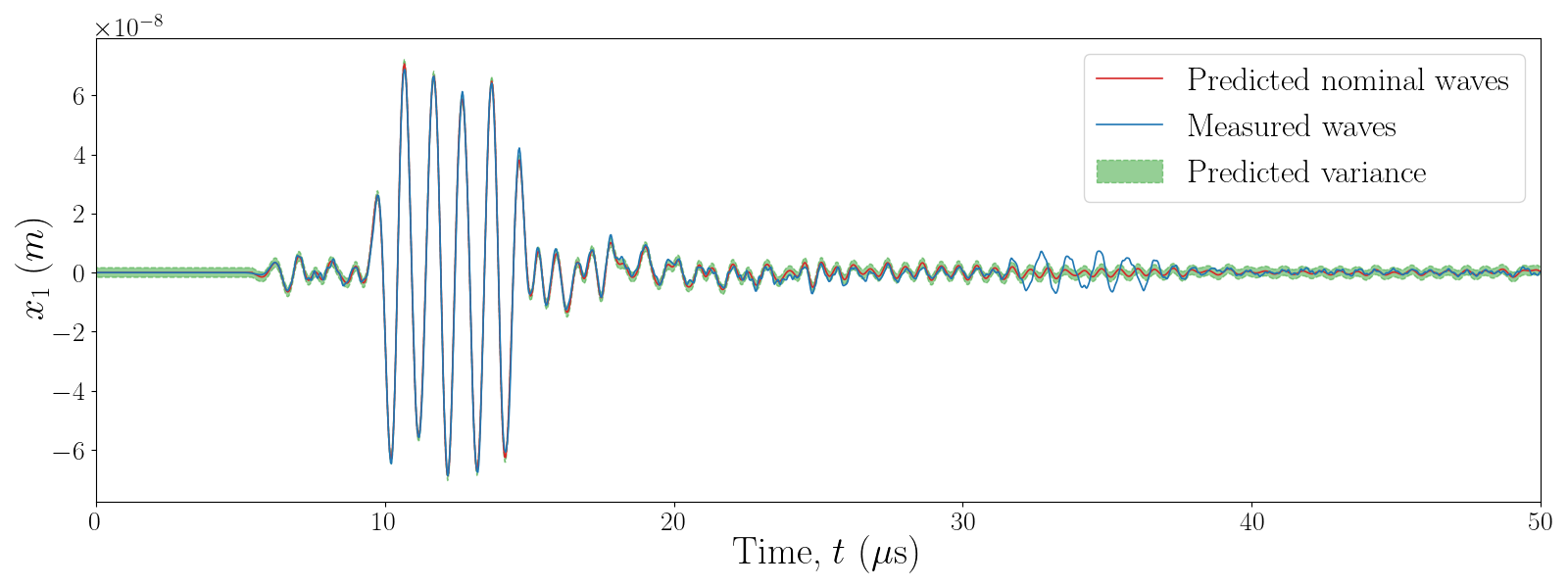}
		\caption{}
		\label{fig:ss_decomp_a-90_d50}
	\end{subfigure}
	\begin{subfigure}[h]{0.95\textwidth}
		\includegraphics[width=\textwidth]{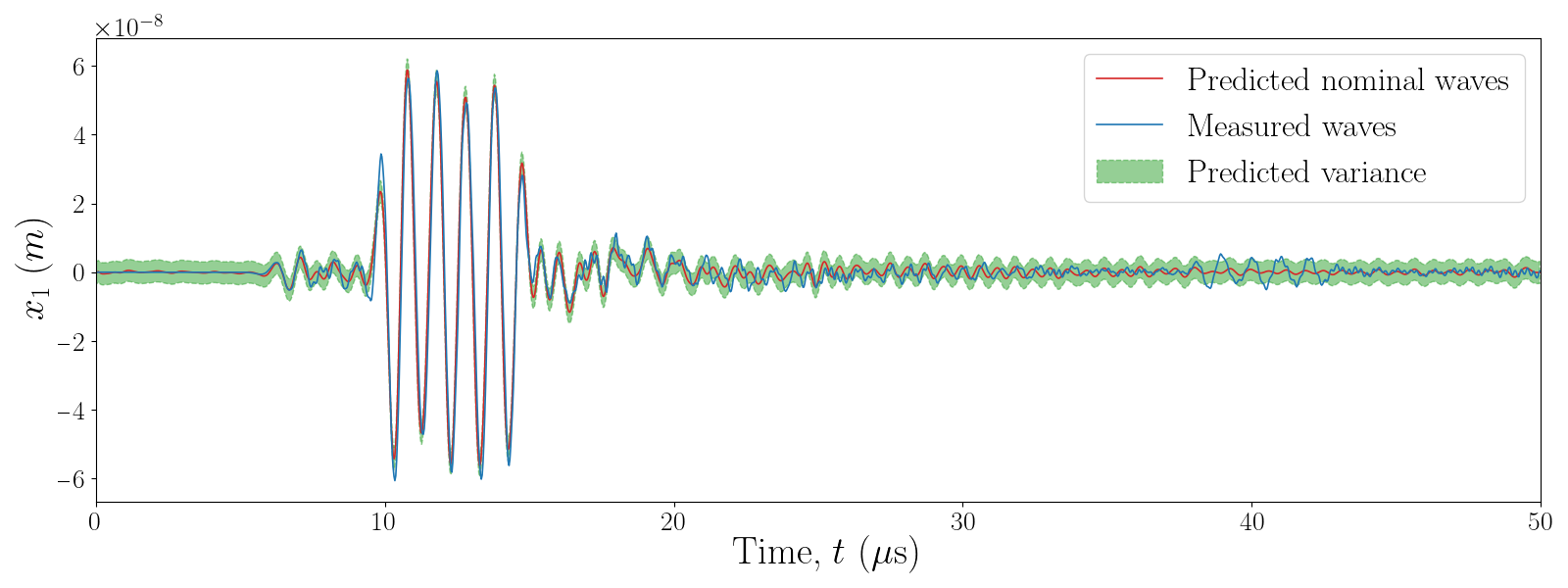}
		\caption{}
		\label{fig:ss_decomp_a-45_d50}
	\end{subfigure}
	\caption{Results from the single-source decomposition applied to signals at a propagation distance of 50mm and propagation angles of (a) $\text{-}90^{\circ}$ and (b)$\text{-}45^{\circ}$. The blue line shows the measured signal, the red line shows the predicted nominal waves and the green boundaries show the expected variance.}
	\label{fig:ss_decomp}
\end{figure}

\Cref{fig:dict_align_phase} shows an example of the nominal dictionary wave $\phi(x)$ used to decompose the measured signal at a propagation distance and angle of 50 \si{mm} and -45\si{\degree} respectively. %
The phase misalignment between the dictionary and measured nominal waves can be seen, which causes the increased predicted variance and decreased likelihood when compared to the same propagation distance at -90\si{\degree}. %
By knowledge of simple geometry and analysis of the grid orientation in \Cref{fig:spatial_sampling}, a propagation angle of -45\si{\degree} will offer the `worst case scenario' in spatial sampling misalignment. %
Even at a maximum spatial sampling misalignment, the probabilistic decomposition works well to determine the predicted nominal waves, as shown in \Cref{fig:ss_decomp_a-45_d50}. %

Results of the decomposition of signals containing reflected waves, at different propagation angles, are shown in \Cref{fig:ss_decomp}, including their predicted variance. %
It appears in \Cref{fig:ss_decomp_a-90_d50} that the signal taken at a location collinear with the actuator and damage contains a reflection signal of larger amplitude. %
Although both signals contain reflected waves, and those in the signal taken at $\text{-}90^{\circ}$ appear stronger, there is a larger predicted variance for the signal taken at $\text{-}45^{\circ}$. %

In \Cref{fig:ss_decomp_a-45_d50}, the reflection because of damage can be seen at approx.\ $36 \leq t \leq 43\mu s$, and is clearly separable from the predicted nominal wave signal. %
There also appears to be a small phase difference between the measured signal and predicted nominal signals for each mode. %
This is a result of the dictionary-measured wave misalignment as shown in \Cref{fig:dict_align_phase}. %
At a similar distance, the measured and predicted nominal waves appear to be a similar shape and amplitude, with the onset and amplitude of the reflected wave being the key difference. %

An important contribution of the probabilistic decomposition used here is the predicted variance of the signal, which gives an indication to the level of uncertainty of the decomposed signal. %
Here, the uncertainty can be used to discuss how confident one is in the decomposed signals at different propagation angles. %
At the highest level of uncertainty -- a propagation angle of 45\si{\degree} -- the confidence in the signal is still relatively large, and so the signal can reasonably be used to determine the reflection source.  %

The initial step of the \emph{in-situ} stages of the methodology shown here is the decomposition of the single-source signal to obtain information of reflected wave signals. %
The decomposition strategy employed here works well and appears to provide the reflection signal needed for accurate localisation. %
\Cref{tab:ss_decomp_metrics} and \Cref{fig:ss_decomp} demonstrated that there is increased uncertainty when decomposing signals at propagation distances different to that used for the initial full-field decomposition. %
As these are waves of short wavelength, small differences in propagation distance can result in significant phase differences between the nominal modes due to their dispersive characteristics. %
In this case, this spatial sampling misalignment did not result in a phase difference large enough to produce an overly-large predicted variance. 
As will be shown in \Cref{sec:local_results}, the dictionary misalignment did not strongly affect the localisation accuracy. %
In all cases, but particularly where spatial sampling is limited, or noise in the signal is comparatively larger -- such as with SLDV methods -- it is important to consider the wavelength with respect to the spatial-sampling step size. %
Subject to further work, the ability to capture the predicted variance of the nominal signals is a key characteristic of this method, as this can be used to better determine any `outlying' signals by only taking the difference when the measured signal exceeds the expected variance. %
Overall, the probabilistic approach works well to decompose the signal into selected modes, doing well to determine the nominal modes and predicted variance. %
The parameters resulting from the method follow what would be expected from the physics, showing the method can also produce ready-to-use features that are individual to each mode. %


\section{Damage Localisation}
\label{sec:local_results}

\subsection{Damage localisation along propagation path}
\label{sec:one_dim_results}

\begin{figure}[h!]
	\includegraphics[width=0.95\textwidth]{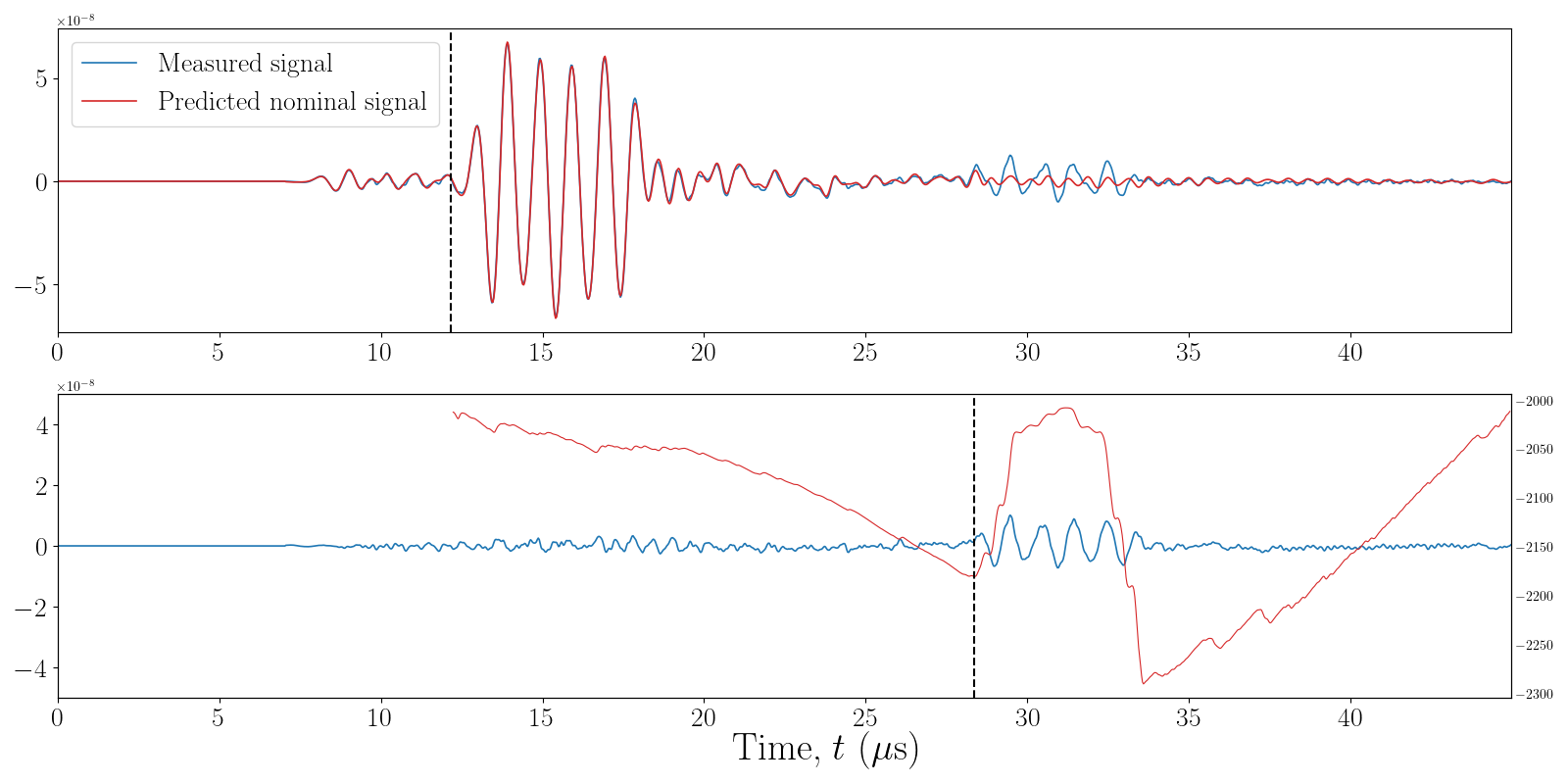}
	\caption{Results of the reflection onset detection stage of signal processing. The upper graph shows the measured and predicted nominal signals, and the lower graph shows the residual signal in blue and the AIC value in red, with the detected onset indicated by the black dashed line. The signal shown is for a propagation distance of 55 \si{mm} at an angle of $\text{-}90^{\circ}$.}
	\label{fig:1d_locate}
\end{figure}

For illustrative purposes, and to assess the onset detection stage on the overall methodology, a one-dimensional localisation strategy along the propagation path containing the damage is shown here. %
An example of the signal processing steps applied from a single source can be seen in \Cref{fig:1d_locate}, showing the decomposed signal, reflected signal and onset determined for the incident $A_0$ mode and reflected $A_0$ mode. %
For both the incident and reflected waves, the onset is determined by the AIC method, and for the reflection, only the portion of the signal after the incident wave arrival is used. %
The residual signal clearly shows the reflection signal, which is much stronger than noise present during the period in the signal containing the nominal waves. %
During the portion of the residual signal in which the nominal waves arrive, there is still some structure as a result of errors between the predicted signal and measured signal. %
As this structure would likely cause a drop in the calculated ratio of information, the AIC (\Cref{eq:AIC}) is calculated for the residual signal beginning from the onset of the slowest incident wave, which is in this case the $A_0$ mode. %
By inspection of the simulation results in \Cref{fig:frame_undam_35} and the residual signal in \Cref{fig:1d_locate}, it can be seen that the reflection signal observed is that of the reflected $A_0$ mode, as the $S_0$ mode reflection is of too little amplitude in comparison to the noise floor. %

The detected onsets align well with the incident and reflection waves seen in the signals. %
There is also a drop in the value of the AIC at the end of the reflected wave, indicating a disparity in the entropy level before and after this point. %
The authors theorise that this is likely because of there being no strong waves appearing in the signal after this point, therefore the signal will be of lower structure than the structured reflection signal. %

The one-dimensional localisation technique outlined in \Cref{sec:localise_method} was applied to signals at various distances along the propagation path. %
Results of the estimated reflection source difference are shown in \Cref{tab:1d_localisation}, as well as their error with respect to the true value. %
There does not appear to be a correlation between distance between the reflection source and sensing signal location, as the error values sharply increase between 10 and 20 mm but proceed to increase slightly at increased distances. %

\begin{table}[h!]
	\centering
	\begin{tabular}{l | l | l | l | l}
		Signal source $\{x,y\}$ & True Distance & Estimated Distance & Error & \% Error \\
		\hline
		$\{-65,0\}$ & 10.0 \si{mm} & 9.910 \si{mm}  & 0.090 \si{mm} & 0.90\%\\
		$\{-55,0\}$ & 20.0 \si{mm} & 20.728 \si{mm} & 0.728 \si{mm} & 3.64\%\\
		$\{-45,0\}$ & 30.0 \si{mm} & 30.827 \si{mm} & 0.827 \si{mm} & 2.76\%\\
		$\{-35,0\}$ & 40.0 \si{mm} & 40.975 \si{mm} & 0.975 \si{mm} & 2.44\%\\
	\end{tabular}
	\caption{Results of damage location estimation along propagation path containing the damage, showing the true distance and estimated distance of the location of the signal and the damage.}
	\label{tab:1d_localisation}
\end{table}

\subsection{Two-dimensional damage localisation}
\label{sec:two_dim_results}

By using the method outlined in \Cref{sec:localise_method}, an estimated location of the damage on the surface of the plate was calculated for various sensor groups. %
The results of these prediction are shown in \Cref{tab:loc_results}, where the error is calculated as the Euclidean distance between the estimated $\hat{E}$ and true location of damage. %
As the reflected wave that is present in the signal is the $A_0$ mode, because of its larger energy than the $S_0$ mode, the group velocity of the $A_0$ mode was used for triangulation. %
From DISPERSE \cite{Pavlakovic1997disperse}, a group velocity for the $A_0$ wave at 1MHz was calculated to be 3071 \si{m/s}. %
The predicted variance of the decomposed nominal waves is also shown in \Cref{tab:loc_results}, and there do not appear to be any significant differences in this variance between sensors. %
Sensor groups with larger error do not appear to contain significantly-larger predicted variances than groups with low prediction error; indicating that any inaccuracies are not a result of the decomposition. %
Furthermore, all the predicted variances are greater than those for the decomposed waves in an undamaged plate (\Cref{fig:BLR_sigma_undam}). %
As the predicted variances are all low in comparison to the signal amplitude, there is confidence in the predicted nominal signal matching the nominal waves. %
This confidence is extended to the confidence of the residual signal containing only reflected/scattered waves. %

\begin{table}[h!]
	\centering
		\begin{tabular}{c|ccc|ccc|cc}
		\multirow{ 2}{*}{Group No.} & \multicolumn{3}{c|}{Sensors} & \multicolumn{3}{c|}{Signal Variance ($\times 10^9$)} & \multicolumn{2}{c}{Predicted Damage Location} \\
		 & \multicolumn{1}{c}{$S_1$} & \multicolumn{1}{c}{$S_2$} & \multicolumn{1}{c|}{$S_3$} & \multicolumn{1}{c}{$\sigma^2_1$} & \multicolumn{1}{c}{$\sigma^2_2$} & \multicolumn{1}{c|}{$\sigma^2_3$} & \multicolumn{1}{c}{$\hat{E} = \{x,y\}$} & \multicolumn{1}{c}{Error (\si{mm})} \\
		\hline
		1 & A & B & C & 1.745 & 3.252 & 2.656 & \{-76.037, 2.37e-15\} & 1.037 \\
		2 & B & C & D & 3.252 & 2.656 & 3.095 & \{-75.252, 0.351\} & 0.4324 \\
		3 & C & D & E & 2.656 & 3.095 & 2.614 & \{-57.965, 2.997\} & \textbf{17.297} \\
		4 & A & C & F & 1.745 & 2.656 & 3.980 & \{-76.037, 2.23e-15\} & 1.037 \\
		5 & B & D & G & 3.252 & 3.095 & 2.138 & \{-68.058, -0.916\} & \textbf{7.002} \\
		6 & C & D & H & 2.656 & 3.095 & 3.981 & \{-75.698, 0.318\} & 0.7671 \\
		\end{tabular}
		\caption{Table of estimated location for various sensor pairings, and the error defined as the Euclidean distance between estimated and true location. The predicted variance of the decomposed signal is also shown for each sensor used. The two sensor groups with particularly high errors (3 \& 5) are highlighted in bold.}
		\label{tab:loc_results}
\end{table}

\begin{figure}[h!]
	\centering
	\begin{subfigure}{0.48\textwidth}
		\includegraphics[width=\textwidth]{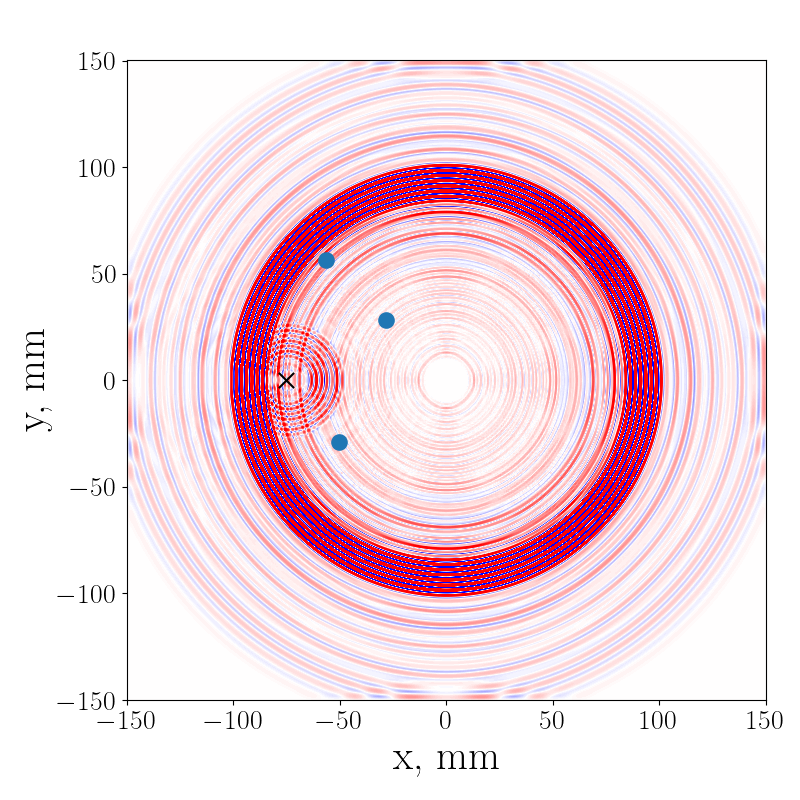}
		\caption{}
	\end{subfigure}
	\begin{subfigure}{0.48\textwidth}
		\includegraphics[width=\textwidth]{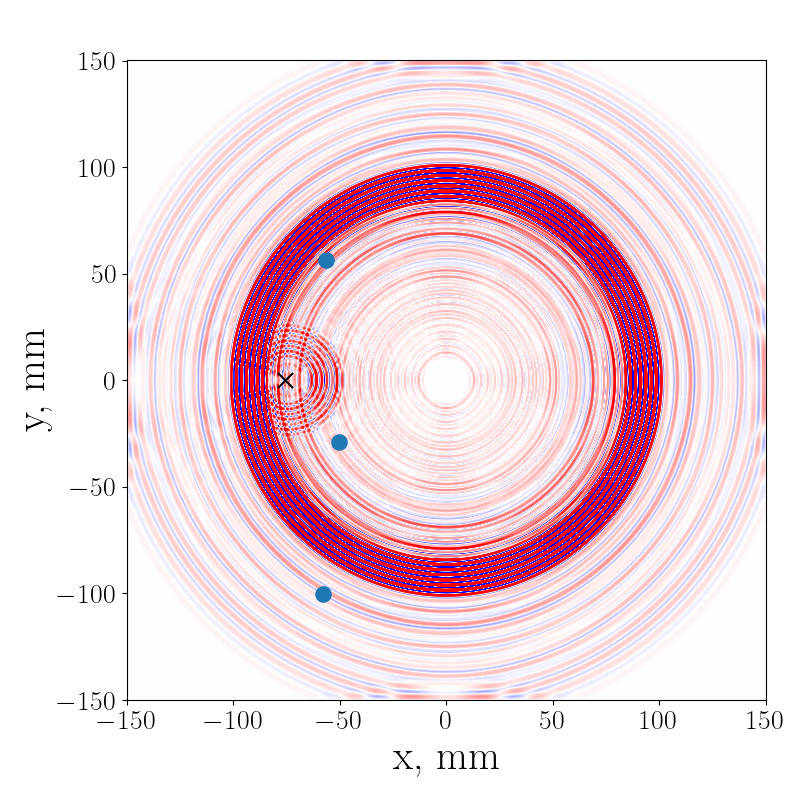}
		\caption{}
	\end{subfigure}
	\caption{Simulation of damaged plate at $35\mu s$, the black cross shows the predicted location of damage for (a) sensor group 2 and (b) sensor group 6. The blue circles indicate the locations of the sensors.}
	\label{fig:pred_location_good}
\end{figure}

A visual representation of predicted locations for the sensor groups with the smallest error are shown in \Cref{fig:pred_location_good}. %
These figures show how accurate the predicted locations were in regards to the whole plate. %
It appears that relative orientation and distance of the sensors from the damage do not influence the accuracy of the method. %
As there appears to be no trend relating inaccuracy to sensor positions, some insight will be made here into the results of the two most inaccurate sensor groups (3 and 5). %
With regards to these groups with the largest error in location (groups 3 and 5), the confidence in the predicted signal is not significantly larger than those with small error. %
This indicates that the source of the error is not from the decomposition of the signal, but from other sources; the following paragraph aims to explore the cause of the error, by looking at the wavefield data at the predicted time-of-arrival of the reflected wave. %

\begin{figure}[h!]
	\centering
	\begin{subfigure}{0.48\textwidth}
		\includegraphics[width=\textwidth]{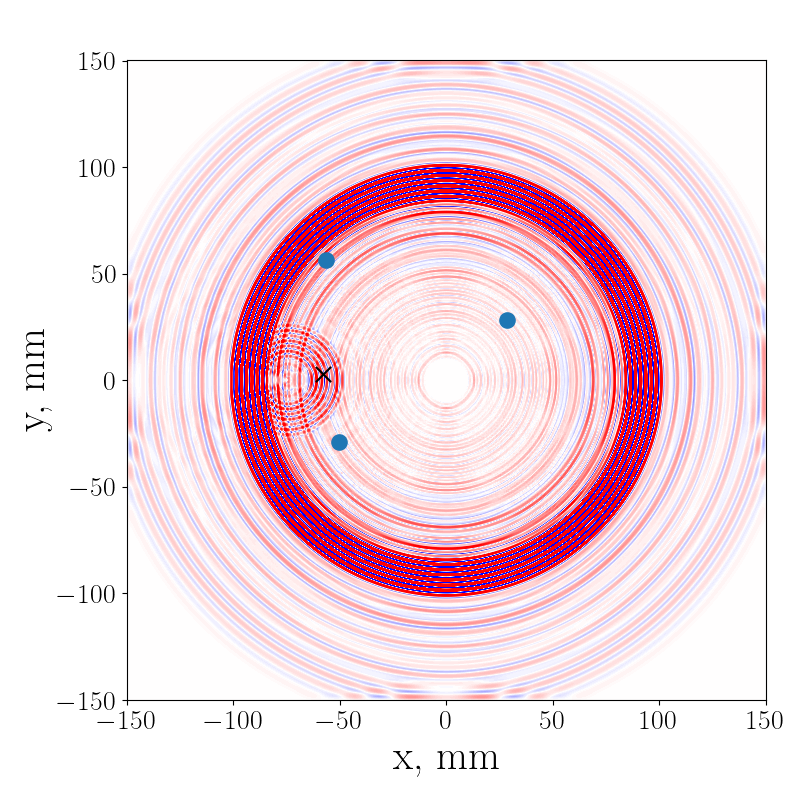}
		\caption{}
		\label{fig:pred_location_grp3}
	\end{subfigure}
	\begin{subfigure}{0.48\textwidth}
		\includegraphics[width=\textwidth]{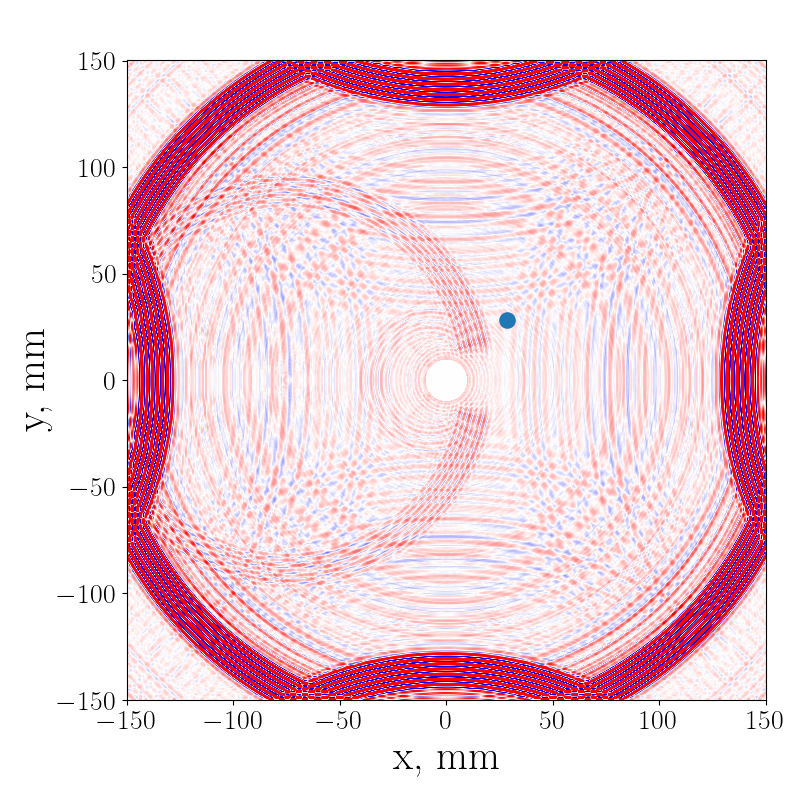}
		\caption{}
		\label{fig:sensE_ref_onset}
	\end{subfigure}
	\caption{(a) Predicated location of damage from sensor group 3 superimposed on the surface displacement at $35\mu s$ and (b) surface displacement at $51.34\mu s$, which corresponds to the calculated TOA for the reflected wave at sensor E; the sensor location is indicated by the blue circle.}
	\label{}
\end{figure}

\Cref{fig:pred_location_grp3} shows the predicted location from the sensor group with the largest error. %
Sensor group 3 includes sensor E, which is a large distance from the damage source. 
However, there appear to be several reflected waves within the signal, and it is difficult to know which of these additional waves is reflected from the damage. %
The surface displacement at the time of the detected reflection onset for sensor E is shown in \Cref{fig:sensE_ref_onset}. %
The detected onset appears to be the time of arrival of the $S_0$ wave reflected from the boundary, as this sensor is at \ang{45} the boundary reflections from the upper and right sides of the plate arrive simultaneously. %
These boundary reflections arrive before the damage reflections and so it is difficult to determine the correct onset without significant prior knowledge. %
For plates of larger size with respect to the distance between sensor and damage, boundary reflections would not be as much of an issue in determining the correct reflection onset as boundary reflections would arrive at the sensor later in time than reflections from damage. %

\begin{figure}[h!]
	\centering
	\begin{subfigure}{0.48\textwidth}
		\includegraphics[width=\textwidth]{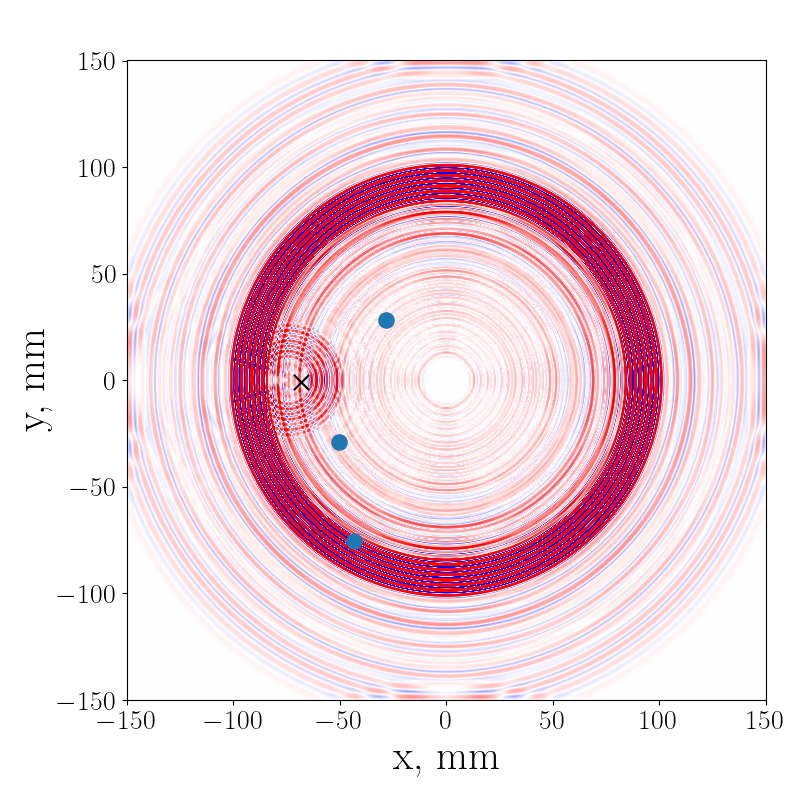}
		\caption{}
		\label{fig:pred_location_grp5}
	\end{subfigure}
	\begin{subfigure}{0.48\textwidth}
		\includegraphics[width=\textwidth]{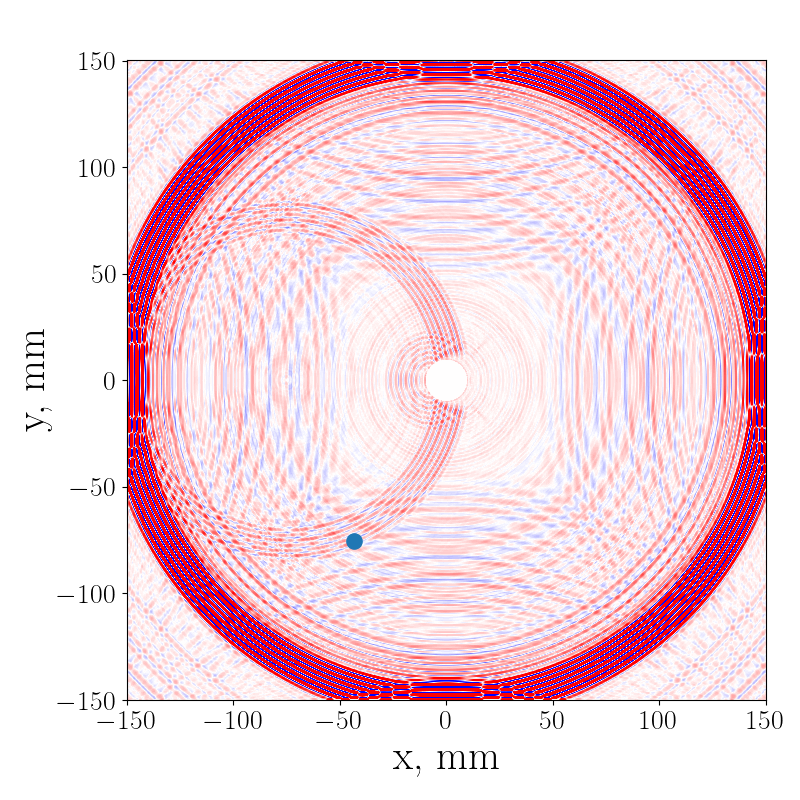}
		\caption{}
		\label{fig:sensG_ref_onset}
	\end{subfigure}
	\caption{(a) Predicated location of damage from sensor group 5 superimposed on the surface displacement at $35\mu s$ and (b) surface displacement at $47.48\mu s$, which corresponds to the calculated TOA for the reflected wave at sensor G; the sensor location is indicated by the blue circle.}
	\label{}
\end{figure}

The predicted location from the sensor group with second largest error is shown in \Cref{fig:pred_location_grp5}, as well as the reflection onset detected from the measured signal. %
As sensor G is not in any of the groups that predicted the location with lower error, it is useful to look at the surface displacement at the time of reflection arrival at this sensor; this is shown in \Cref{fig:sensG_ref_onset}. %
At this time point, there does not appear to be any other reflection arriving at the sensor; in fact, the damage reflection does appear to be arriving at this time. %
However, from visual inspection of \Cref{fig:sensG_ref_onset}, it appears that the true arrival time of the damage reflection was earlier. %
The inaccuracy in the onset detection -- much like with sensor group 3 -- may be explained by the large number of incident/and reflected waves that are in the signal at the same time as the damage reflection. %

\subsection{Discussion}
\label{sec:discussion}

The work in this paper is intended as a proof-of-concept of the probabilistic decomposition of guided waves, analysed with an objective of locating inhomogeneities in plates. %
The localisation results using the decomposition method demonstrated here shows strong capability as a proof-of-concept, in this case study locating damage accurately to within 1 \si{mm}. %
By using a variety of sensor placements and groups to triangulate, it appears that sensor group orientation does not influence accuracy of location. %
Errors in localisation from sensor placement is due to the influence of boundary reflections. %
This independence is expected as the plate was modelled with a homogeneous material and accurate wave speeds. %
This can also be attributed to the quality of the decomposition remaining high at different propagation angles, as shown in \Cref{sec:decomp_results}. %

The main aim of the paper here was to demonstrate the unique aspects of the probabilistic decomposition, which has some advantages and promising attributes. %
The probabilistic decomposition method contributes well to a damage detection strategy, as the returned parameters can be used to assess the deconstructed signal. %
All the single-source signals were decomposed with a reasonable level of confidence, shown by the relatively low $\sigma^2$ parameter. %
The ability to assess the confidence in nominal wave decomposition is particularly important for multi-regime strategies; i.e.\ using SLDV data for the NWD and PZT sensors for single-source. 
Furthermore, there is no need for prior knowledge of the material properties for decomposition at any stage, making the method highly adaptable and applicable to scenarios where material properties are unknown. %

The strongest influence on inaccurate predicted location of damage was reflections of the incident waves from the plate boundary and actuator. %
The plate simulated here was relatively small (limited by computational resources), which is useful in order to determine dispersion curves as reflection signals are necessary to capture this information. %
However, for the purposes of damage detection, this size plate is not likely to be inspected and so the influence of boundary reflections would be minimal. %
In situations where reflections not caused by damage are expected, such as complex geometries or fasteners, so long as these reflections are included in the data fed into the 2DFT algorithm, either by accurate modelling or by experimental capture, the nominal wave dictionary will include these additional signals. %

As stated in \Cref{sec:LISA}, the simulated damage size was chosen to allow reasonably large amplitude of the reflection signal, but still have low enough amplitude to be smaller value than the nominal waves. %
This was done in order to test the capability of the method on relatively weak reflection signals, it should be noted that reflections from the boundary are going to be of larger amplitude thanks to the higher reflection coefficient. %
A single damage type was chosen here (half-thickness notch), which also influences the reflective characteristics of the inhomogeneity. %

As previously stated, the strongest influence on inaccurately-predicted locations was that of boundary reflections. %
The decomposition stage determines the additional waves well, but determining the onset using the AIC method can become more difficult when there are many structured signals superimposed on top of each other. %
This also highlights the need for consideration of sensor locations close to reflective boundaries, however, this is a consistent problem in UGW-based localisation methods, and not distinct to this methodology \cite{Kundu2014}. %
In this paper, sensors were chosen relatively close to the damage, due to the small size of the plate resulting in quick superposition of many waves, and this is a proof-of-concept study. %
In practical applications, the location of the sensors will need to be carefully considered with respect to any complex geometrical features, or other reflective characteristics of the system. %
As well as sensor configuration, in practical application, the effect of environmental conditions would be important to consider, particularly in fibre-composite materials. %
In fact, with knowledge of the effect of these factors, the dimensionality of the nominal wave dictionary could be further extended to include such features. %

There are many proposed methods for determining the time of arrival of the first wave in a signal, such as wavelet transforms, analysis of the long-time or short-time average (LTA/STA) and higher-order statistics \cite{Lokajicek2006}.
Another potential method of mitigating the influence of boundary reflections is, much like with complex geometries, to increase the fidelity of the nominal-wave dictionary to include \emph{expected} reflections. %
In fact, when measuring signals at the propagation angle used in the full-field decomposition stage, the reflected signals are included in the reconstructed individual modes. %
Therefore, one could mitigate boundary reflection influence by including propagation angle as an alignment feature of the nominal-wave dictionary; however, this would significantly increase computation and data storage costs. %
The propagation angle influence on nominal waves can be enhanced with machine learning tools, by utilising models such as those shown in \cite{Haywood2021}. %

An advantage of the method shown in this paper is the applicability to more complex materials and structures because its only prerequisite information is surface displacement data along the propagation path, rather than an accurate analytical mode of the waveforms. %
Furthermore, as complex materials can introduce nonlinearity \cite{kundu2018nonlinear}, this method will directly incorporate the signal shape due to nonlinearity without the need for a complex physical model. %
The Bayesian approach to decomposition is another benefit, as it determines a distribution of predicted models rather than just the best fit. %
This uncertainty can be utilised to assess the predicted signal and therefore confidence in the residual signal containing only reflected waves. %
As well as this, the method is efficient in terms of computation, memory and storage, and can be done with low-cost sensors. %
Overall, the Bayesian tools applied to decompose single-source signals for localisation allow for additional metrics that can aid in assessment of calculated residual signals. %
These probabilistic measures can aid in applications such as active learning \cite{bull2019probabilistic}. %

For practical applications, this method could be used across multiple stages of a pitch-catch guided-wave damage detection system, as the decomposition method inherently produces features which can be used to detect the existence of (and potentially characterise) damage. %
This decomposition stage then also directly contributes to any localisation procedure, as decomposition of guided waves is a standard method to determine reflection signals. %
The main contribution of the method is that the decomposition not only produces inherent parametric features, but also gives a measure of the confidence in the decomposition. %
Both of these characteristics indicate this method to be useful for many NDE and SHM strategies, which often use probabilistic approaches throughout. %

The technology presented here could be applied to non-destructive evaluation, as well as system identification, in the aerospace and wind energy industry. Furthermore, the technology presented here can inform traditional, well-established tools, such as data-based vibration analysis in structural health monitoring, as well as modal analysis. Also, they can form the basis on expanding on guided wave experimental work, for example that by Flynn, et al. \cite{flynn2011maximum}, Haywood-Alexander, et al. \cite{Haywood2021}, or Willberg, et al. \cite{Willberg2012}. As a final point, the technology presented here can be generalised for physics-based modelling, using a machine learning probabilistic framework, in particular for signal decomposition, and potentially system identification on finite element modelling.

\subsection{Future work}
For the localisation strategy demonstrated here, some prior knowledge was used; i.e., the known group velocity and knowledge that the reflected wave seen in the residual signal is that of the $A_0$ mode, because of the stronger reflection characteristics of this mode at this wavelength and damage size. %
The purpose of this work was to demonstrate the localisation strategy based on decomposition; however, it is important to discuss using this with no prior knowledge. %
The authors intend to do further work into several scenarios when prior knowledge is not available, for reasons such as system complexity or lack of information. %
There may be scenarios where the wave speed is uncertain; i.e., in anisotropic materials where the wave speed is dependent on propagation angle, or when the wave detected in the residual signal is unknown as to which reflected mode it is. %
When the wave mode is unknown, a number of mode detection strategies will be tested, including: using dispersion curve information to determine the temporal order of arrival of modes, improving the localisation optimisation \Cref{eq:local_opt} to include dTOA between reflection and nominal waves, and matching the strength and group velocity between reflection and nominal signals. %
When the wave speed is unknown, it is more difficult to determine the temporal order of modes in the signal. %
In this case, the authors intend to utilise a probabilistic approach such as that found in \cite{jones2022bayesian}. %
Probabilistic approaches lend themselves to damage-detection strategies via their inherent uncertainty metrics which can be propagated through to uncertainty in location. %
Therefore, using a probabilistic approach in both the decomposition and localisation stages of the methodology may be advantageous. %

The authors also intend to extend the dimensionality of the nominal wave dictionary to include other parameters which have been discussed in this paper to affect the nominal wave signal. %
Some examples of additional features include; propagation angle, environmental factors, actuation frequency. %

\section{Conclusion}
A methodology for localisation of damage using ultrasonic guided waves has been demonstrated here by simulating Lamb waves in an aluminium plate. %
The overall strategy involves decomposing measured single-source signals, from which residual signals can be determined which contain reflected waves, which are then used to triangulate the reflection source. %
The decomposition stage uses a Bayesian approach in order to generate a distribution of possible nominal waves, allowing better determination of uncertainty. %
In particular, the full-field decomposition strategy has potential for use in systems where accurate physical models of signals are difficult to obtain. %
In this paper, the accuracy of the decomposition and localisation methods have been shown. %
Inaccurate location was found to be largely because of the influence of boundary reflections, which is not likely to be an issue in application as the plate shown here was of small size. %

Discussion has been made here of the importance of prior knowledge on the localisation strategy, and how to improve the methodology if this information was not known. %
Further work will be done to employ a probabilistic approach to the localisation stage when wave speed is unknown. %
In particular, for the cases of anisotropic materials causing propagation-angle-dependent velocities and unknown wave mode conversion. %

\section{Acknowledgements}
The authors gratefully acknowledge the support of the UK Engineering and Physical Sciences Research Council (EPSRC) [grant numbers EP/R004900/1, EP/R003645/1 and EP/N010884/1].

\begin{appendices}
\crefalias{section}{appendix}

\section{Bayesian Linear Regression}
\label{app:BLR}
\setcounter{equation}{0}
\renewcommand{\theequation}{\thesection.\arabic{equation}}

Traditional linear regression formulates a model using point estimates of a set of parameters which "best" fit an available dataset, based on minimising an $L^2$-norm between the model predictions and the data. %
Instead, BLR aims to establish a probability distribution of possible model parameters. %
The model has the form, 
\begin{equation}
    y = \mathbf{w}^{\top} \phi(\mathbf{x}) + \varepsilon, \qquad \varepsilon \sim \mathcal{N} (0,\sigma^2)
\end{equation}
where $\phi$ is some basis for expansion of a $p$-dimensional data point $\mathbf{x}$; $\mathbf{w} = \{ w_1,w_2, ... ,w_p \}$ are the associated weights of the basis expansion, and $\varepsilon$ is an additive Gaussian white noise distributed as $\mathcal{N}(0,\sigma^2)$. %
The weights $\mathbf{w}$ and the variance $\sigma^2$ are the unknowns. %
The Bayesian linear regression model approach was chosen since it returns a quantified uncertainty. 
The task is then to compute the posterior distribution of the parameters $p(\mathbf{w},\sigma^2|D)$. %
This posterior distribution has the form,
\begin{equation}
	p(\mathbf{w},\sigma^2|\mathcal{D}) = NIG(\mathbf{w},\sigma^2|\mathbf{w}_N,\mathbf{V}_N,a_N,b_N)
\end{equation}
with,
\begin{equation}
	\mathbf{w}_N = \mathbf{V}_N(\mathbf{V}_0^{-1}\mathbf{w}_0+\mathbf{X}^{\top}\mathbf{y})
	\label{eq:post_weights}
\end{equation}
\begin{equation}
	\mathbf{V}_N = (\mathbf{V}_0^{-1}+\mathbf{X}^{\top}\mathbf{X})^{-1}
	\label{eq:post_var}
\end{equation}
\begin{equation}
	a_N = a_0 +n/2
	\label{eq:post_a}
\end{equation}
\begin{equation}
	b_N = b_0 +\frac{1}{2}\left(\mathbf{w}_0^{\top} \mathbf{V}_0^{-1} \mathbf{w}_0 + \mathbf{y}^{\top}\mathbf{y} - \mathbf{w}_N^{\top} \mathbf{V}_N^{-1}\mathbf{w}_N \right)
	\label{eq:post_b}
\end{equation}
where $\mathbf{V}_0$, $\mathbf{w}_0$, $a_0$ and $b_0$ are hyperparameters of the prior. %
It is possible to set a less-informative prior for $\sigma^2$ by applying $a_0=b_0=0$. %
Also setting $\mathbf{w}_0 = 0$ and $\mathbf{V}_0 = g(\mathbf{X}^{\top}\mathbf{X})^{-1}$ for any positive value $g$; leads to Zellner's \textit{g-prior} \cite{Zellner1986}. %
By having the prior variance proportional to $(\mathbf{X}^{\top}\mathbf{X})^{-1}$, it is ensured that the posterior is invariant under scaling of the inputs. %

An important metric that is attainable from the Bayesian linear regression method is the \emph{predictive likelihood}, which gives an indication of the likelihood that the model fits and takes into account the uncertainty as well as the quality of the mean fit. %
As this method uses a tractable Gaussian posterior, the predictive likelihood is given by, 
\begin{equation}
	p(\tilde{\mathbf{y}}|\tilde{\mathbf{x}},\mathbf{y},\mathbf{x}) = \mathcal{N}\left(\mathbb{E}\left[\tilde{\mathbf{y}}\right],\mathbb{V}\left[\tilde{\mathbf{y}}\right]\right)
\end{equation}
where $\mathbb{E}\left[\tilde{\mathbf{y}}\right]$ and $\mathbb{V}\left[\tilde{\mathbf{y}}\right]$ are the predicted values and variance of the output given input $\mathbf{x}$. %
For computational stability this likelihood is calculated in the log space, and is named the \emph{independent predictive log-likelihood} $PLL_i$ and is defined by,
\begin{equation}
	PLL_i = \sum_{i}^{N}\log \mathcal{N}(\mathbf{y}_{i}\vert\mathbb{E}[\mathbf{y}_{i}],\mathbb{V}[\mathbf{y}_{i}],\mathbf{w})
\end{equation}
for $N$ data points. This value is the product over the predictive likelihoods for every point, i.e.\ the joint likelihood if they were uncorrelated. %

\begin{algorithm}[ht]
	\caption{\textsl{Bayesian Linear Regression.}}
	  \label{alg:BLR}
	  \SetAlgoLined
	  \SetKwInOut{Input}{Input}\SetKwInOut{Output}{Output}
	  \Input{~~Design matrix input data $\mathbf{X}$, target data $\mathbf{y}$, predictive regressors $\tilde{\mathbf{X}}$, prior hyperparameters $\mathbf{w}_0,\; \mathbf{V}_0,\; a_0,\; b_0$}
	  \Output{~~Predicted model $\tilde{\mathbf{y}}$, predicted variance of model $\sigma^2$}
	\BlankLine
	\BlankLine
	  \nosemic \textbf{Posterior distribution}\;
	  \textit{Calculate} posterior variance $\mathbf{V}_N$ using \cref{eq:post_var}\;
	  \textit{Calculate} posterior weights $\mathbf{w}_N$ using \cref{eq:post_weights}\;
	  \textit{Calculate} posterior hyperparameters $a_N,\; b_N$ using \cref{eq:post_a,eq:post_b}\;
	  \nosemic \textbf{Posterior predictive}\;
	  \nosemic \textit{Calculate} posterior predicted mean:\;
	  \pushline \dosemic $\tilde{\mathbf{y}} = \mathbf{w}_N^{\top}\mathbf{X}$\;
	  \popline \nosemic \textit{Calculate} posterior predicted variance:\;
      \pushline \dosemic $\mathbb{V} = \frac{b_N}{a_N}(\mathbb{I}_m + \tilde{\mathbf{X}}\mathbf{V}_N\tilde{\mathbf{X}}^{\top})$\;
	  \popline \nosemic \textit{Return} model predicted variance:\;
	  \pushline \dosemic $\sigma^2 = \textrm{diag}(\mathbb{V})$\;
\end{algorithm}

			  

\end{appendices}

\bibliography{biblio}


\end{document}